\documentclass[12pt,epsfig]{article}
\usepackage[usenames]{color}
\usepackage{epsfig}

\newcommand{\be}{\begin{eqnarray}}
\newcommand{\ee}{\end{eqnarray}}

\newcommand{\beq}{\begin{equation}}
\newcommand{\eeq}{\end{equation}}
\newcommand{\lam}{\lambda}
\newcommand{\vk}{{\bf k}}
\newcommand{\nn}{\nonumber}

\def\la{\mathrel{\mathpalette\fun <}}
\def\ga{\mathrel{\mathpalette\fun >}}
\def\fun#1#2{\lower3.6pt\vbox{\baselineskip0pt\lineskip.9pt
\ialign{$\mathsurround=0pt#1\hfil##\hfil$\crcr#2\crcr\sim\crcr}}}

\begin{document}

\newcommand{\bea}{\begin{eqnarray}}
\newcommand{\eea}{\end{eqnarray}}
\def\la{\mathrel{\mathpalette\fun <}}
\def\ga{\mathrel{\mathpalette\fun >}}
\def\fun#1#2{\lower3.6pt\vbox{\baselineskip0pt\lineskip.9pt
\ialign{$\mathsurround=0pt#1\hfil##\hfil$\crcr#2\crcr\sim\crcr}}}
\def\wh{\widehat}
\newcommand{\gmp}{g_\mu^\top}
\newcommand{\gnp}{g_\nu^\top}
\newcommand{\gmn}{g_{\mu\nu}}
\newcommand{\gmnp}{g_{\mu\nu}^\top}
\newcommand{\gmnpp}{g_{\mu\nu}^{\top\top}}
\newcommand{\gabpp}{g_{\alpha\beta}^{\top\top}}
\newcommand{\gmmpri}{g_{\mu\mu'}}
\newcommand{\gmmprip}{g_{\mu\mu'}^{\top}}
\newcommand{\gnnprip}{g_{\nu\nu'}^{\top}}
\newcommand{\gmmpripp}{g_{\mu\mu'}^{\top\top}}
\newcommand{\gnnpripp}{g_{\nu\nu'}^{\top\top}}
\newcommand{\mfi}{m_\phi}
\newcommand{\bk}{{\bf k}}
\newcommand{\gampp}{\gamma_\mu^{\top\top}}
\newcommand{\ganpp}{\gamma_\nu^{\top\top}}
\newcommand{\kompp}{k_{1\mu}^{\top\top}}
\newcommand{\ktmpp}{k_{2\mu}^{\top\top}}
\newcommand{\konpp}{k_{1\nu}^{\top\top}}
\newcommand{\kpp}{k^2_{\top\top}}
\newcommand{\kompripp}{k^{'\top\top}_{1\mu}}
\newcommand{\konpripp}{k^{'\top\top}_{1\nu}}
\newcommand{\koapp}{k^{\top\top}_{1\alpha}}
\newcommand{\kobpp}{k^{\top\top}_{1\beta}}
\newcommand{\aaa}{\alpha}
\newcommand{\dd}{\delta}
\newcommand{\vx}{{\bf x}}
\newcommand{\vy}{{\bf y}}
\newcommand{\vp}{{\bf p}}
\newcommand{\vq}{{\rm {\bf q}}}
\newcommand{\vF}{{\bf F}}
\newcommand{\vE}{{\bf E}}
\newcommand{\vB}{{\bf B}}
\newcommand{\mT}{{\rm\bf T}}
\newcommand{\mS}{{\rm\bf S}}
\newcommand{\vV}{{\bf V}}
\newcommand{\tilphi}{\tilde{\varphi}}
\newcommand{\ka}{\mbox{\rm \ae}}
\newcommand{\hm}{\hspace{-1.5ex}}
\newcommand{\ep}{\epsilon}
\newcommand{\pp}{\partial}
\newcommand{\bp}{\mbox{\boldmath$p$}}
\newcommand{\bg}{\mbox{\boldmath$\gamma$}}
\newcommand{\bs}{\mbox{\boldmath$\sigma$}}
\newcommand{\bta}{\mbox{\boldmath$\tau$}}
\newcommand{\bq}{\mbox{\boldmath$q$}}
\newcommand{\bb}{\mbox{\boldmath$b$}}
\newcommand{\al}{\alpha}
\newcommand{\bet}{\beta}
\newcommand{\bkp}{\bf k_\perp}

\newcommand{\KL}{\rm\Lambda K^+}
\newcommand{\KS}{\rm\Sigma K}
\newcommand{\er}{$\pm$}
\newcommand{\widt}{\rm\Gamma_{tot}}
\newcommand{\wadd}{$\rm\Gamma_{miss}$}
\newcommand{\gpiN}{$\rm\Gamma_{\pi N}$}
\newcommand{\getN}{$\rm\Gamma_{\eta N}$}
\newcommand{\gkla}{$\rm\Gamma_{K \Lambda}$}
\newcommand{\gksi}{$\rm\Gamma_{K \Sigma}$}
\newcommand{\gNpi}{$\rm\Gamma_{P_{11} \pi}$}
\newcommand{\gDpf}{$\rm\Gamma_{\Delta\pi(L\!<\!J)}$}
\newcommand{\gDps}{$\rm\Gamma_{\Delta\pi(L\!>\!J)}$}
\newcommand{\sqgDpf}{$\rm\sqrt\Gamma_{\Delta\pi(L\!<\!J)}$}
\newcommand{\sqgDps}{$\rm\sqrt\Gamma_{\Delta\pi(L\!>\!J)}$}
\newcommand{\gNpf}{$\rm\Gamma_{D_{13}\pi}$}
\newcommand{\gNps}{$\Gamma_{D_{13}\pi(L\!>\!J)}$}
\newcommand{\gnsi}{$\rm N\sigma$}
\newcommand{\roper}{$ N(1440)P_{11}$}
\newcommand{\srma}{$  N(1535)S_{11}$}
\newcommand{\trma}{$ N(1520)D_{13}$}
\newcommand{\srmb}{$ N(1650)S_{11}$}
\newcommand{\trmb}{$ N(1700)D_{13}$}
\newcommand{\trmc}{$ N(1875)D_{13}$}
\newcommand{\trmd}{$ N(2170)D_{13}$}
\newcommand{\fvma}{$ N(1675)D_{15}$}
\newcommand{\fvmb}{$ N(2070)D_{15}$}
\newcommand{\fvpa}{$ N(1680)F_{15}$}
\newcommand{\srpb}{$ N(1710)P_{11}$}
\newcommand{\trpa}{$ N(1720)P_{13}$}
\newcommand{\trpb}{$ N(2200)P_{13}$}
\newcommand{\trpd}{$ N(2170)D_{13}$}
\newcommand{\dtpa}{$\Delta(1232)P_{33}$}
\newcommand{\doma}{$\Delta(1620)S_{31}$}
\newcommand{\dtma}{$\Delta(1700)D_{33}$}
\newcommand{\dtmb}{$\Delta(1940)D_{33}$}
\newcommand{\bc}{\begin{center}}
\newcommand{\ec}{\end{center}}

\title{
The analysis of  reactions $\pi N\to two\, mesons + N$ within reggeon exchanges.\\
2. Basic formulas for fit.
}

\author{V.V. Anisovich and A.V. Sarantsev \\
{\it Petersburg Nuclear Physics Institute, Gatchina, 188300, Russia}}

\maketitle

\begin{abstract}
We present technical aspects of the
fitting procedure given in the paper V.V. Anisovich and A.V. Sarantsev
{\it The analysis of  reactions $\pi N\to two\, mesons + N$ within reggeon exchanges.
1. Fit and results.}
\end{abstract}

 \vspace{0.5cm}


\section{Introduction}

The novel point of the analysis
given in the paper \cite{km08}
is a direct use of reggeon exchange
technique for the description of the reactions
$\pi N\to two\, mesons + N$ at
large energies of the initial pion. This approach allows us to
describe simultaneously distributions over $M$ (invariant
mass of two mesons) and $t$ (momentum transfer squared to nucleons).
Making use of this technique, the following resonances (as well as
corresponding bare states), produced in the
$\pi N\to \pi^0\pi^0 N$
reaction were studied: $f_0(980)$, $f_0(1300)$, $f_0(1200-1600)$,
$f_0(1500)$, $f_0(1750)$, $f_2(1270)$, $f_2(1525)$, $f_2(1565)$,
$f_2(2020)$, $f_4(2025)$.
Adding the data on the processes
  $p\bar p({\rm at\, rest,\, from\, liquid\,H_2})\to \pi^0\pi^0\pi^0$,
 $\pi^0\pi^0\eta$, $\pi^0\eta\eta$
 and
$p\bar p({\rm at\, rest,\, from\, gaseous\, H_2})\to \pi^0\pi^0\pi^0$,
$\pi^0\pi^0\eta$, $\pi^0\eta\eta$,
we
performed simultaneous $K$-matrix fit of two-meson spectra in all
these reactions. In \cite{km08}, the results of the combined fit to the above-listed
isoscalar $f_J$-states and isovector ones,
$a_0(980)$, $a_0(1475)$, $a_2(1320)$, were presented.

Here we discuss  technical aspects of the
fitting procedure used in \cite{km08}. To give the guide   to
a reader through the states, which are discussed in \cite{km08},
 the assignment of mesons to Regge trajectories as well as to
$q\bar q$ nonets is presented following Ref. \cite{book3}.

The paper is organized as follows.

In Section 2, we  present the necessary elements of the
reggeon exchange technique in the two-meson
production reactions.
In Section 3, we discuss the status of the Regge trajectories on
$(J,M^2)$-plane.
In Section 4, we give the connection between the  dispersion integral
equation for a three-body system
and the $K$-matrix approach.
We demonstrate what type of assumptions is needed
to transform the three-body dispersion relation amplitudes
into those of the $K$-matrix approach.
The assignment of mesons to
$q\bar q$ nonets is given in Section 5.

 \section{Elements of the
reggeon exchange technique in the two-meson
production reactions}

Here we present the details for the partial wave
analysis of two-meson system produced in the high energy $\pi N$ interaction when
the two-meson production occurs due to reggeon exchanges.
The reggeon exchange approach is a good tool for studying hadron
binary reactions and processes with diffractive production of hadrons
at high energies (see \cite{book2}, Chapters 2 and 6).
Interference effects in the amplitudes of the type
$\pi N\to two\, mesons + N$ provide  valuable
information on the contributions of  resonances with different
quantum numbers.

\subsubsection{Kinematics for reggeon exchange amplitudes}

For illustration, we consider the reaction $\pi^- p \to \pi\pi+n$ in
the c.m. system of the reaction and present the momenta of the
incoming and outgoing particles (below we use the notation for
four-vectors: $x=(x_0,
\vec x_\top , x_z$).

For the incoming particles we have:
\bea
\label{Tk1}
{\rm pion\, momentum}&:&\quad
p_1=(p_z+\frac{m^2_\pi}{2p_z},0,p_z)\, , \\
{\rm proton\, momentum}&:&\quad
p_2=(p_z+\frac{m^2_N}{2p_z},0,-p_z)\, , \nn \\
{\rm total \, energy\, squared}&:&\quad s_{\pi N}=(p_1+p_2)^2\, . \nn
\eea
Here we have performed an expansion over the large momentum $p_z$.
Analogously, we write for the outgoing particles:
 \bea
\label{Tk2}
&&{\rm meson \,momenta\, } (i=1,2):\quad k_i= (k_{iz}-\frac{m^2_i
+k^2_{i\top}}{2k_{iz}},\vec k_{i\top},k_{iz}
)\, , \\
&&{\rm total\, momentum\, of\, mesons}:\; P=k_1+k_2=
p_1-q=(p_z+\frac{s+m^2_\pi
+2q^2_\top}{4p_z},\vec q_\top,p_z-
\frac{s-m^2_\pi}{4p_z})\, ,\nn \\
&&{\rm proton \,momentum}:\quad k_3=p_2+q=(p_z-\frac{s-m^2_\pi
+2q^2_\top}{4p_z},-\vec q_\top,-p_z+
\frac{s-m^2_\pi}{4p_z})\, ,\nn \\
&&{\rm energy\, squared\, of\,
 mesons }:\quad s=P^2=(k_1+k_2)^2\, .
\eea \label{Tk2a}

The relative momenta of mesons in the initial and final states read:
\be
  p=\frac 12(p_1+q),\quad k=\frac 12(k_1-k_2).
\ee
 The momentum squared transferred to the nucleon is
comparatively small:\\ $t\equiv q^2 \sim m^2_N << s_{\pi N}$ where
\be \label{Tk3}
q= ( -\frac{s+m^2_\pi +2q^2_\top}{4p_z},-\vec q_\top,
\frac{s-m^2_\pi}{4p_z})\, .
\ee
Neglecting $0(1/p^2_z)$-terms, one has
$q\simeq ( 0,-\vec q_\top, 0)$ and
$q^2 \simeq -q^2_\top $.

\subsubsection{ Angular momentum operators for two-meson systems}

As in Ref. \cite{km08}, we use angular momentum operators
$X^{(L)}_{\mu_1\ldots\mu_L}(k^\perp)$,
$Z^{\alpha}_{\mu_1\ldots\mu_L}(k^\perp)$ and projection operator
$O^{\mu_1\ldots\mu_L}_{\nu_1\ldots\nu_L}(\perp P)$. Recall their
definition.

The operators are constructed from the relative momenta
$k^\perp_\mu$ and tensor $g^\perp_{\mu\nu}$. Both of them are
orthogonal to the total momentum of the system:
\be
k^\perp_\mu=\frac12
g^\perp_{\mu\nu}(k_1-k_2)_\nu =k_{1\nu} g^{\perp P}_{\nu\mu}
=-k_{2\nu} g^{\perp P}_{\nu\mu}\ ,  \qquad
g^\perp_{\mu\nu}=g_{\mu\nu}-\frac{P_\mu P_\nu}{s}\;.
\ee

The operator for $L=0$ is a scalar (we write $X^{(0)}(k^\perp)=1$),
and the operator for $L=1$ is a vector, $X^{(1)}_\mu=k^\perp_\mu $.
The operators $X^{(L)}_{\mu_1\ldots\mu_L}$ for $L\ge 1$ can be written
in the form of a recurrency relation:
\bea
X^{(L)}_{\mu_1\ldots\mu_L}(k^\perp)&=&k^\perp_\alpha
Z^{\alpha}_{\mu_1\ldots\mu_L}(k^\perp)\equiv  k^\perp_\alpha
Z_{\mu_1\ldots\mu_L,\alpha}(k^\perp) \; ,
\nonumber\\
Z^{\alpha}_{\mu_1\ldots\mu_L}(k^\perp)&\equiv &
Z_{\mu_1\ldots\mu_L,\alpha}(k^\perp)=
\frac{2L-1}{L^2}\Big (
\sum^L_{i=1}X^{{(L-1)}}_{\mu_1\ldots\mu_{i-1}\mu_{i+1}\ldots\mu_L}(k^\perp)
g^\perp_{\mu_i\alpha}
\nonumber \\
 -\frac{2}{2L-1}  \sum^L_{i,j=1 \atop i<j}
&g^\perp_{\mu_i\mu_j}&
X^{{(L-1)}}_{\mu_1\ldots\mu_{i-1}\mu_{i+1}\ldots\mu_{j-1}\mu_{j+1}
\ldots\mu_L\alpha}(k^\perp) \Big ).
\label{Vz}
\eea
We have a convolution equality
$X^{(L)}_{\mu_1\ldots\mu_L}(k^\perp)k^\perp_{\mu_L}=k^2_\perp
X^{(L-1)}_{\mu_1\ldots\mu_{L-1}}(k^\perp)$,
with $k^2_\perp\equiv k^\perp_{\mu}k^\perp_{\mu}$, and
the tracelessness property of $X^{(L)}_{\mu\mu\mu_1\ldots\mu_L}=0$.
On this basis,
one can write down the orthogonali\-ty--normalization condition for
orbital angular  operators:
\be
\int\frac{d\Omega}{4\pi}
X^{(L)}_{\mu_1\ldots\mu_L'\ldots\mu_L}(k^\perp)X^{(L')}_{\mu_1\ldots\mu_L'}(k^\perp)
 = \delta_{LL'}\alpha_L k^{2L}_\perp \; ,\quad
\alpha_L\ =\ \prod^L_{l=1}\frac{2l-1}{l}  ,
\label{Valpha}
\eea
where integration is performed over spherical variables: $\int d\Omega/(4\pi)=1$.

Iterating equation (\ref{Vz}), one obtains the
following expression for the operator $X^{(L)}_{\mu_1\ldots\mu_L}$:
\bea
\label{Vx-direct}
&&X^{(L)}_{\mu_1\ldots\mu_L}(k^\perp)=
\alpha_L \bigg [
k^\perp_{\mu_1}k^\perp_{\mu_2}k^\perp_{\mu_3}k^\perp_{\mu_4}
\ldots k^\perp_{\mu_L} \\
&&-\frac{k^2_\perp}{2L-1}\bigg(
g^\perp_{\mu_1\mu_2}k^\perp_{\mu_3}k^\perp_{\mu_4}\ldots
k^\perp_{\mu_L}
+g^\perp_{\mu_1\mu_3}k^\perp_{\mu_2}k^\perp_{\mu_4}\ldots
k^\perp_{\mu_L} + \ldots \bigg)
\nn \\
&&+\frac{k^4_\perp}{(2L\!-\!1)(2L\!-\!3)}\bigg(
g^\perp_{\mu_1\mu_2}g^\perp_{\mu_3\mu_4}k^\perp_{\mu_5}
k^\perp_{\mu_6}\ldots k^\perp_{\mu_L}
+
g^\perp_{\mu_1\mu_2}g^\perp_{\mu_3\mu_5}k^\perp_{\mu_4}
k^\perp_{\mu_6}\ldots k^\perp_{\mu_L}+
\ldots\bigg)+\ldots\bigg ].\nn
\eea
For the projection operators, one has:
\be
&&\hspace{-6mm}O= 1\ ,\qquad O^\mu_\nu (\perp P)=g_{\mu\nu}^\perp \, ,\nn \\
&&\hspace{-6mm}O^{\mu_1\mu_2}_{\nu_1\nu_2}(\perp P)=
\frac 12 \left (
g_{\mu_1\nu_1}^\perp  g_{\mu_2\nu_2}^\perp \!+\!
g_{\mu_1\nu_2}^\perp  g_{\mu_2\nu_1}^\perp  \!- \!\frac 23
g_{\mu_1\mu_2}^\perp  g_{\nu_1\nu_2}^\perp \right ).\;\;
\ee
For higher states, the operator can be calculated using the
recurrent expression:
\be
&&O^{\mu_1\ldots\mu_L}_{\nu_1\ldots\nu_L}(\perp P)=
\frac{1}{L^2} \bigg (
\sum\limits_{i,j=1}^{L}g^\perp_{\mu_i\nu_j}
O^{\mu_1\ldots\mu_{i-1}\mu_{i+1}\ldots\mu_L}_{\nu_1\ldots
\nu_{j-1}\nu_{j+1}\ldots\nu_L}(\perp P)
 \\
&&- \frac{4}{(2L-1)(2L-3)}
\times
\sum\limits_{i<j\atop k<m}^{L}
g^\perp_{\mu_i\mu_j}g^\perp_{\nu_k\nu_m}
O^{\mu_1\ldots\mu_{i-1}\mu_{i+1}\ldots\mu_{j-1}\mu_{j+1}\ldots\mu_L}_
{\nu_1\ldots\nu_{k-1}\nu_{k+1}\ldots\nu_{m-1}\nu_{m+1}\ldots\nu_L}(\perp P)
\bigg ). \nn
\ee
The projection operators
 obey the
relations:
\bea
O^{\mu_1\ldots\mu_J}_{\nu_1\ldots\nu_J}(\perp P)
X^{(J)}_{\nu_1\ldots\nu_J}(k^\perp)&=&
X^{(J)}_{\mu_1\ldots\mu_J}(k^\perp)\, ,\nn \\
O^{\mu_1\ldots\mu_J}_{\nu_1\ldots\nu_J}(\perp P) k_{\nu_1} k_{\nu_2}
\ldots k_{\nu_J} &=& \frac
{1}{\alpha_J}X^{(J)}_{\mu_1\ldots\mu_J}(k^\perp)\; .
\eea
 Hence, the product of the two $X^J(k_\perp)$ operators results in the Legendre
polynomials as follows:
\be
X^{(J)(k_\perp)}_{\mu_1\ldots\mu_J}(p^\perp) (-1)^J
O^{\mu_1\ldots\mu_J}_{\nu_1\ldots\nu_J}(\perp P)
X^{(J)}_{\nu_1\ldots\nu_J}(k^\perp)\!=\!\alpha_J
(\sqrt{-p_\perp^2}\sqrt{-k_\perp^2})^J P_J(z),
\ee
where $z\equiv (-p^\perp k^\perp)/(
\sqrt{-p_\perp^2}\sqrt{-k_\perp^2})$
.

\subsection{ Reggeized pion exchanges}

To be definite,
we present here formulas which lead to differential cross-section
moment expansion in processes initiated by  reggeized pions situated on
the trajectories $R(\pi_j)$.
Recall that index $j$ labels different trajectories, which are
leading and daughter ones,
as well as trajectories generated by a reggeization of the $t$-channel states with $J>0$
(in multiparticle processes, like $\pi N\to {two\, mesons}+N$, the vertices $\pi R(\pi_j)$
 are diffrent for states with $J=0$ and $J>0$, so it is convenient to separate their
 contributions).

 \subsubsection{ Production of two mesons at small $|t|$ -- the hypothesis
 of dominant pion exchange}

Under this hypothesis, the
amplitude for the $\pi\pi$ production block is written as follows:
\bea
&&A(\pi R(\pi_j)\to\pi\pi)\to A(\pi\pi\to\pi\pi)=16\pi \sum\limits_{J} A^J_{\pi
\pi\to\pi\pi}(s) (2J\!+\!1) N^0_JY^0_J(z,\varphi), \\
&&
Y^m_J(z,\varphi)=\frac{1}{N^m_J}P^m_J(z)e^{im\varphi}, \quad
N^m_{J}=\sqrt{\frac{4\pi}{2J\!+\!1}\frac{(J+m)!}{(J-m)!}}.\nn
\eea
Below we consider the  resonance decay in the set of channels
with two pseudoscalar mesons in the final states, that means
$\pi\pi\to c=\pi\pi, K\bar K,\eta\eta,\ldots$  Generalizing consideration,
we give the same freedom for initial state.
Therefore, we
denote the initial and final $(I=0)$-channel states as $a,c$ with
 $a,c=\pi\pi, K\bar K,\eta\eta$, and so on. Then the
transition amplitudes are denoted as
\be X^{(J)}_{\mu_1\ldots\mu_J}(p^\perp) \, A^J_{a\to
c}(s)(-1)^J O^{\mu_1\ldots\mu_J}_{\nu_1\ldots\nu_J} (\perp P)
X^{(J)}_{\nu_1\ldots\nu_J}(k^\perp_c)\xi_J \, , \qquad
\xi_J=\frac{16\pi(2J+1)}{\alpha_J},
\ee
where $k^\perp_{c\mu}=\frac12(k^\perp_{c1\nu}-k^\perp_{c2\nu})g^\perp_{\nu\mu}$.

The unitarity
condition for the transition amplitudes reads:
\be
{\rm Im} A^J_{a\to c}(s)= \sum\limits_b
\frac{2\sqrt{-k_{b\perp}^2}}{\sqrt{s}} A^J_{a\to b}(s) A^{J*}_{b\to
c}(s)(-k_{c\perp}^2)^J\,,
\label{Tunit_1}
\ee
where index $b$ refers to intermediate states ($b=\pi\pi$, $K\bar K$,
$\eta\eta,\ldots$).
The unitarity
condition is fulfilled by using for $A^J_{a\to c}(s)$
the $K$-matrix form:
\be
A^J_{a\to c}(s)= \sum\limits_b \hat K^J_{ab}
\bigg(\frac{I}{I-i\hat \rho^J(s) \hat K^J}\bigg)_{bc}\ ,
\ee
where $\hat \rho$ is a diagonal matrix with elements
$\rho^J_{bb}(s)=$ $2\sqrt{-k_{b\perp}^2} (-k_{b\perp}^2)^J$ $/ \sqrt s$.

We parametrize the elements of the $K$-matrix in the following form:
\be
\label{T6-a-B-1}
K^J_{ab}&=&\sum\limits_n \frac{1}{B_J(-k_{a\perp}^2,r_n)}
\left (\frac{g^{n(J)}_a g^{n(J)}_b}{M_n^2-s}\right
)\frac{1}{B_J(-k_{b\perp}^2,r_n)} \nn \\&+&
\frac{f^{(J)}_{ab}}{B_J(-k_{a\perp}^2,r_0)B_J(-k_{b\perp}^2,r_0)}\ .
\ee
Index $n$ refers to a set of resonances,
the resonance couplings  $g^n_c$ are
constants, and $f_{ac}$ is a non-resonance term. The
form factors $B_J(-k_\perp^2,r)$ are introduced to compensate the
divergence of the relative momentum factor at large energies. Such
 form factors are known as the Blatt--Weisskopf factors depending on
the radius of the state $r_n$. For non-resonance transition, the
radius $r_0$ is taken to be much larger than that for resonance
contributions.

\subsubsection{Calculation routine for the reggeized pion exchanges}

In  case of the two-meson production, the initial-state $K$-matrix
element is called the $P$-vector:
$K^J_{\pi R(\pi_j), b}\equiv P^J_{\pi R(\pi_j), b}$
(recall that  index $j$ labels the leading and daughter trajectories).
We write for the two-meson production amplitude initiated by the pion
reggeon exchange the following representation:
\be
A^J_{\pi R(\pi_j), c}(s)=\sum\limits_b P^J_{\pi R(\pi_j), b}
\bigg(\frac{I}{I-i\hat \rho^J(s) \hat K^J}\bigg)_{bc} \, .
\ee
 The $P$-vector is parametrized in the form similar to eq. (\ref{T6-a-B-1}):
\be
P^J_{\pi R(\pi_j), c}&=&\sum\limits_n
\frac{1}{B_J(-p_\perp^2,r_n)} \left (\frac{G^{n(J)}_{\pi R(\pi_j)}(q^2)
g^{n(J)}_c}{M_n^2-s}\right
)\frac{1}{B_J(-k_{c\perp}^2,r_n)} \nn \\&+&
\frac{F^{(J)}_{\pi R(\pi_j),c}(q^2)}{B_J(-p_\perp^2,r_0)B_J(-k_{c\perp}^2,r_0)}\, .
\ee
The product of the two amplitudes is equal to:
\bea
&&\hspace{-9mm}A(\pi
R(\pi_j)\!\to\!\pi\pi)A^*(\pi R(\pi_k)\!\to\!\pi\pi)\!= (16\pi)^2
\sum\limits_J Y^0_J(z,\varphi) \nn \\
&&\hspace{-10mm} \times\sum\limits_{J_1 J_2} d_{J_1J_2 J}^{~0\,0\,0}
A^{J_1}_{\pi R(\pi_j)\to \pi\pi}(s)A^{J_2}_{\pi R(\pi_k)\to
\pi\pi}(s)(2J_1\!+\!1)(2J_2\!+\!1)N^0_{J_1} N^0_{J_2} \, ,
\eea
where the coefficients $d_{nmk}^{iji+j}$ are given below.
 Averaging over the polarizations of the initial nucleons and summing
over polariation of the final ones, we get $ {\rm Sp} [(\vec
\sigma \vec q_\perp)(\vec \sigma \vec q_\perp)] \simeq -q^2=-t\, $.
So, we obtain for the total amplitude squared:
\bea \label{Tamp_2}
 |A_{\pi p\to\pi\pi n}^{({\rm pion\, trajectories})}|^2 &=&
\sum\limits_{R(\pi_j)R(\pi_k)}
A(\pi R(\pi_j)\to\pi\pi)A^*(\pi R(\pi_k)\to\pi\pi)\nn \\
&\times& R_{\pi_j}(s_{\pi N},q^2) R^*_{\pi_k}(s_{\pi N},q^2)(-t)
(g^{(\pi)}_{pn})^2 .
\eea
The final expression reads:
\bea
&&N(M,t){\langle{Y}^0_J\rangle}=\frac{\rho(s)\sqrt s}{\pi|\vec
p_2|^2s_{\pi N}} \sum\limits_{R(\pi_j)R(\pi_k)} R_{\pi_j}(s_{\pi
N},q^2) R^*_{\pi_k}(s_{\pi N},q^2)(-t) (g^{(\pi)}_{pn})^2 \nn \\
&&\times\sum\limits_{J_1 J_2} d_{J_1J_2 J}^{~0\,0\,0}
A^{J_1}_{\pi\pi_j\to \pi\pi}(s)A^{J_2}_{\pi R(\pi_k)\to
\pi\pi}(s)(2J_1\!+\!1)(2J_2\!+\!1)N^0_{J_1} N^0_{J_2} .
\eea

{\centerline{\bf  (i)
 Spherical functions. }}

Let us present here some relations for the spherical functions used
in the calculations:
\bea
&&Y^m_l(\Theta, \varphi)=\frac{1}{N^m_{l}}P^m_l(z)e^{im\varphi},
\quad N^m_{l}=\sqrt{\frac{4\pi}{2l+1}\frac{(n+m)!}{(l-m)!}}\,, \nn \\
&& P^m_l(z)=(-1)^m(1-z^2)^{\frac{m}{2}}\frac{d^m}{dz^m}P_l(z),
\eea
where $z=\cos \Theta$. We have the following convolution rule for
 two spherical functions:
\be
Y^i_n(\Theta, \varphi)Y^j_m(\Theta, \varphi)=
\sum_{k=0}^{n+m}d^{i,j,i+j}_{n,m,k}Y^{i+j}_k(\Theta, \varphi)\ ,
\ee
$d^{i,j,i+j}_{n,m,k}$ is defined also in \cite{km08},  eq. (8).

\subsubsection {Calculations related to the expansion of the
differential cross section $\pi p \to \pi\pi +N$ over spherical
functions for the reggeised $\pi_2$-exchange}

Here we present formulas which refer to the calculation routine
related to the reggeised $\pi_2$-exchange.

The convolution of angular momentum operators can be expressed
through Legendre polynomials and their derivatives:
\bea
&& X^{(J+2)}_{\alpha\beta\mu_1\ldots\mu_J}(p^\perp)(-1)^J
O^{\mu_1\ldots\mu_J}_{\nu_1\ldots\nu_J}(\perp P)
X^{(J)}_{\nu_1\ldots\nu_J}(k^\perp)
\\
&&=\frac{2\alpha_J
\left (\sqrt{-k^2_\perp}\right )^J \left (\sqrt{-p^2_\perp}\right
)^{J+2}}
{3(J\!+\!1)(J\!+\!2)} \nn
\\
&&\times\left(
X^{(2)}_{\mu\nu}(p^\perp)\frac{P''_{J+2}}{-p^2_\perp}+
X^{(2)}_{\mu\nu}(k^\perp)\frac{P''_{J}}{-k^2_\perp}-
\frac{3\,P''_{J+1}}{\sqrt{-k^2_\perp} \sqrt{-p^2_\perp}} k^\perp_\mu
p^\perp_\nu \right )O^{\alpha\beta}_{\mu\nu}(\perp q)\ , \nn
\\
&& O_{\chi\tau}^{\alpha\beta}(\perp q)
X^{(J)}_{\chi\mu_2\ldots\mu_J}(p^\perp)(-1)^J
O^{\tau\mu_2\ldots\mu_J}_{\nu_1\nu_2\ldots\nu_J}(\perp P)
X^{(J)}_{\nu_1\ldots\nu_J}(k^\perp) \nn
\\
&&= \frac{2\alpha_{J-1}}{3J^2} \left (\sqrt{-k^2_\perp}\right )^J
\left (\sqrt{-p^2_\perp}\right )^J \nn
\\
&&\times \left( X^{(2)}_{\mu\nu}(p^\perp)\frac{P''_{J}}{-p^2_\perp}+
X^{(2)}_{\mu\nu}(k^\perp)\frac{P''_{J}}{-k^2_\perp}-
\frac{P'_{J}+2zP''_J}{\sqrt{-k^2_\perp} \sqrt{-p^2_\perp}}
 k^\perp_\mu p^\perp_\nu \right )O_{\mu\nu}^{\alpha\beta}(\perp q)\ ,
\nn \\
&& X^{(J-2)}_{\mu_3\ldots\mu_J}(p^\perp)(-1)^J
O^{\alpha\beta\mu_3\ldots\mu_J}_{\nu_1\nu_2\nu_3\ldots\nu_J}(\perp
P)
X^{(J)}_{\nu_1\ldots\nu_J}(k^\perp)\nn
 \\
&&= \frac{2\alpha_{J-2}\left (\sqrt{-k^2_\perp}\right )^J \left
(\sqrt{-p^2_\perp}\right )^{J-2}} {3(n\!-\!1)n}
\nn
\\
&&\times\left(
X^{(2)}_{\mu\nu}(p^\perp)\frac{P''_{J-2}}{-p^2_\perp}+
X^{(2)}_{\mu\nu}(k^\perp)\frac{P''_J}{-k^2_\perp}-
\frac{3\,P''_{J-1}}{\sqrt{-k^2_\perp} \sqrt{-p^2_\perp}} k^\perp_\mu
p^\perp_\nu \right )O^{\alpha\beta}_{\mu\nu}(\perp q).\nn
\eea
Let us remind that $p_{\mu}^\perp=p_{1\nu}g^{\perp P}_{\nu\mu}$ and
$k_{\mu}^\perp=k_{1\nu}g^{\perp P}_{\nu\mu}$. Therefore the
amplitude $A_{\alpha\beta}(\pi R(\pi_2)\to\pi\pi)$ can be rewritten as:
\bea  \label{Tpi2_xxo}
&&A_{\alpha\beta}(\pi R(\pi_2)\to\pi\pi)= \frac 23
\sum\limits_J\bigg[
\frac{X^{(2)}_{\alpha\beta}(p^\perp)}{-p_\perp^2}\left(
C^{(J)}_1P''_{J+2}A^{(J)}_{+2}(s)+ \right.\bigg . \nn
 \\
&&\left .
+C^{(J)}_2P''_{J}A_0^{(J)}(s)+C^{(J)}_3P''_{J-2}A^{(J)}_{-2}(s)\right
) \nn \\&&+
\frac{X^{(2)}_{\alpha\beta}(k^\perp)}{-k_\perp^2}P''_J\left(
C^{(J)}_1A^{(J)}_{+2}(s)+C^{(J)}_2A_0^{(J)}(s)+
C^{(J)}_3A^{(J)}_{-2}(s)\right )~~
\nn \\
&&-\frac{O^{\alpha\beta}_{\mu\nu}k^\perp_\mu
p^\perp_\nu}{\sqrt{-k^2_\perp} \sqrt{-p^2_\perp}} \left(
3C^{(J)}_1P''_{J+1}A^{(J)}_{+2}(s)+C^{(J)}_2(P'_J+2zP''_J)A^{(J)}_0(s)\right.
\nn \\ &&\left. \bigg . + 3C^{(J)}_3P''_{j-1}A^{(J)}_{-2}(s)\right
)\bigg],
\eea
where
\bea
C^{(J)}_1&=&\frac{16\pi(2J\!+\!1)}{(J\!+\!1)(J\!+\!2)}\left
(\sqrt{-k^2_\perp}\right )^J \left (\sqrt{-p^2_\perp}\right )^{J+2}\
,
\nn \\
C^{(J)}_2&=&\frac{16\pi(2J\!+\!1)}{J(2J\!-\!1)}\left
(\sqrt{-k^2_\perp}\right )^J \left (\sqrt{-p^2_\perp}\right )^{J}\ ,
\nn \\
C^{(J)}_3&=&\frac{16\pi(2J\!+\!1)}{(2J\!-\!1)(2J\!-\!3)}\left
(\sqrt{-k^2_\perp}\right )^J \left (\sqrt{-p^2_\perp}\right )^{J-2}\
.
\eea
In the amplitude with the $X^{(2)}_{\alpha\beta}(p^\perp)$ structure
there is no $m=1$ component. This amplitude should be taken
effectively into account by the $\pi$ trajectory. The second
amplitude has the same angular dependence $P''_J(z)$ and works
 for resonances with $J\ge 2$. In the first approximation
it is reasonable to use  the third term only, which has the smallest
power of $p_\perp^2$.

The third amplitude has angular dependencies:
\be
P''_{J+1}(z)\,,\qquad P'_{J}+2zP''_J\,, \qquad P''_{J-1}\,.
\ee
 The first and second angular dependencies are the same for $J=1,2$
and differ only at $n\ge 3$, when the third term appears. Therefore,
in the first approximation one can use only the second term which
has a lower order of $p_\perp^2$ to fit the data.  Thus,
the $R(\pi_2)$
exchange amplitude can be approximated as:
\be
 A_{\alpha\beta}(\pi R(\pi_2)\to\pi\pi)&\simeq&\frac 23 \sum\limits_J
\left[ \frac{X^{(2)}_{\alpha\beta}(k^\perp)}{-k_\perp^2}P''_J
C^{(J)}_3A^{(J)}_{-2}(s)
\right. \nn \\
&-&\left. \frac{O^{\alpha\beta}_{\mu\nu}k^\perp_\mu
p^\perp_\nu}{\sqrt{-k^2_\perp} \sqrt{-p^2_\perp}}
C^{(J)}_2(P'_J+2zP''_J)A_0^{(J)}(s)\right ].~~~~~~
\label{Tpi2_fin}
\ee
The convolution of operators in (\ref{Tpi2_fin}) with $k^{\perp q}_{3\alpha}
k^{\perp q}_{3\beta}$ in the GJ system gives:
\bea &&
k^{\perp q}_{3\alpha}k^{\perp q}_{3\beta}X^{(2)}_{\alpha\beta} (k^2_\perp)= |\vec
k|^2 k^{\perp q}_{3z} \left(k^{\perp q}_{3z} P_2(z)+3 k_{3x}z\cos\varphi
\sin\Theta\right )\ , \nn \\
&& k^{\perp q}_{3\alpha}k^{\perp q}_{3\beta}O_{\mu\nu}^{\alpha\beta} k^\perp_\mu
p^\perp_\nu= \frac 13|\vec k||\vec p|k^{\perp q}_{3z} \left(2 k^{\perp q}_{3z}
z+3k^{\perp q}_{3x} \cos\varphi \sin\Theta\right )\ ,
\eea
and the total amplitude  is equal to:
\be
\label{Tamp_pi2_tot}
  A_{\pi p\to\pi\pi n}^{R(\pi_2)}&=&\frac{1}{s^2_{\pi N}}\sum\limits_J
 \left (V_1^{J}A^{(J)}_{-2}(s)-V_2^{J}A_{0}^{(J)}\right )
R_{\pi_2}(s_{\pi N},q^2)\nn \\
&\times& \left(\varphi_n^+(\vec\sigma \vec p_\perp)\varphi_p \right)
g^{(\pi_2)}_{pn}\,,
\ee
where
\bea
V_1^{(J)}&=& C^{(J)}_3 k^{\perp q}_{3z} \left(k^{\perp q}_{3z} P_2(z)+3
k_{3x}z\cos\varphi \sin\Theta\right )P''_{J} ,
\nn \\
V_2^{(J)}&=& \frac 13C^{(J)}_2 k^{\perp q}_{3z} \left(P'_{J}+2zP''_J\right )
\left(2 k^{\perp q}_{3z} z+3k^{\perp q}_{3x} \cos\varphi \sin\Theta\right ).
\eea

For $J=1$ the first vertex is equal to 0; for the next ones we have:
\bea
&V_2^{(1)}&=\frac 13C^{(1)}_2 k^{\perp q}_{3z} \left(2
 k^{\perp q}_{3z} Y^0_1N^0_{1}- 3k^{\perp q}_{3x}{\rm Re}\, Y^1_1 N^1_{1} \right )\, ,
 \\
&V_1^{(2)}&= C^{(2)}_3 k^{\perp q}_{3z} \left( k^{\perp q}_{3z} Y^0_2 N^0_{2} -
k_{3x} {\rm Re}\, Y^1_2 N^1_{2} \right ),  \nn
 \\
&V_2^{(2)}&=\frac 13C^{(2)}_2 k^{\perp q}_{3z} \left(12 k^{\perp q}_{3z} Y^0_2
N^0_{2}+6 k^{\perp q}_{3z} Y^0_0 N^0_{0}- 9k^{\perp q}_{3x} {\rm Re}\, Y^1_2 N^1_{2}
\right ),\nn
\\
&V_1^{(3)}&= C^{(3)}_3k^{\perp q}_{3z}\left( 9k^{\perp q}_{3z} Y^0_3
N^0_{3}+18k^{\perp q}_{3z} Y^0_1 N^0_{1} -6k_{3x} Re Y^1_3 N^1_{3} -9k_{3x}
{\rm Re}\, Y^1_1 N^1_{1} \right ),
\nn \\
&V_2^{(3)}&=\frac 13C^{(3)}_2 k^{\perp q}_{3z} \left(30 k^{\perp q}_{3z} Y^0_3
N^0_{3}+42 k^{\perp q}_{3z} Y^0_1 N^0_{1}- 15k^{\perp q}_{3x} {\rm Re}\, Y^1_3
N^1_{3}\right. \nn \\
&&-\left.18k^{\perp q}_{3x} Re Y^1_1 N^1_{1} \right ),\nn \\
&V_1^{(4)}&= C^{(4)}_3 k^{\perp q}_{3z}\left( 18k^{\perp q}_{3z} Y^0_4
N^0_{4}+20k^{\perp q}_{3z} Y^0_2 N^0_{2} +7k^{\perp q}_{3z} Y^0_0 N^0_{0} \right
.\nn
\\
&&-\left. 9k_{3x} {\rm Re}\, Y^1_4 N^1_{4} -15k_{3x} {\rm
Re}\, Y^1_2 N^1_{2} \right ),
\nn
\\
&V_2^{(4)}&=\frac {C^{(4)}_2}{3}k^{\perp q}_{3z}\left [ k^{\perp q}_{3z}\left(56
Y^0_4
N^0_{4}+110 Y^0_2 N^0_{2} +34 Y^0_0 N^0_{0}\right )\right. \nn \\
&&-k^{\perp q}_{3x}\left. \left(21\,{\rm Re}\, Y^1_4 N^1_{4}-30\, {\rm Re}\,
Y^1_2 N^1_{2} \right )\right ]. \nn
\eea
In a general form, the expression can be written as:
\bea
V_1^{(J)}&=& \sum\limits_{n=0}^J  C^{(J)}_3 k^{\perp q}_{3z}\left[ k^{\perp q}_{3z}
Y^0_n R^0_n(P_2P''_J)+3 k^{\perp q}_{3x} {\rm Re}\,\,Y^1_n
R^1_n(zP''_J)\right ],
\nn \\
V_2^{(J)}&=& \sum\limits_{n=0}^J  C^{(J)}_2 k^{\perp q}_{3z}\bigg[\frac 23
k^{\perp q}_{3z} Y^0_n R^0_n(z(P'_J+2zP''_J)) \bigg. \nn \\&+&\bigg.
k^{\perp q}_{3x} {\rm Re}\,\,Y^1_n R^1_n(P'_J+2zP''_J)\bigg ],
\eea
where
\be
R^0_n(f)&=&\int \frac{d\Omega}{4\pi} f(z)Y^0_{n}(z,\Theta),\nn \\
R^1_n(f)&=&2\int \frac{d\Omega}{4\pi} f(z)\cos\varphi \sin\Theta
{\rm Re}\, Y^1_{n}(z,\Theta).
\ee
  The $P$-vector amplitudes for the reggeized $\pi_2$-exchanges read:
\be
A^{(J)}_{-2}(s)=\hat P^{(J)}_{-2} (I-i\hat \rho^J(s) \hat K^J)^{-1},\nn \\
A^{(J)}_0(s)=\hat P^{(J)}_{0} (I-i\hat \rho^J(s) \hat K^J)^{-1}.
\ee
 The $P$-vector components are parametrized in the form:
\bea
\left (P^{(J)}_{-2}\right )_n&=&\sum\limits_\alpha
\frac{1}{B_{J-2}(-p_\perp^2,r_\alpha)} \left
(\frac{G^{(J)\alpha}_{-2} g^{\alpha(J)}_n}{M_\alpha^2-s}\right
)\frac{1}{B_J(-k_{n\perp}^2,r_\alpha)} \nn \\ &+&
\frac{F^{(J)}_{(-2)n}}{B_{J-2}(-p_\perp^2,r_0)B_J(-k_{n\perp}^2,r_0)}
,
\nn \\
\left (P^{(J)}_{(0)}\right )_{n}&=&\sum\limits_\alpha
\frac{1}{B_J(-p_\perp^2,r_\alpha)} \left (\frac{G^{(J)\alpha}_{0}
g^{\alpha(J)}_n}{M_\alpha^2-s}\right
)\frac{1}{B_J(-k_{n\perp}^2,r_\alpha)}\nn \\
&+&
\frac{F^{(J)}_{(0)n}}{B_J(-p_\perp^2,r_0)B_J(-k_{n\perp}^2,r_0)}\,~~~~~~~
\eea
The total amplitude of the $\pi_2$ exchange can be rewritten as an
expansion over spherical functions:
\be
\label{Tamp_pi2_tf}
A_{\pi p\to\pi\pi n}^{(\pi_2)}\!=\!\sum\limits_{n=0}^N \left (Y^0_n
A_{tot}^{0(n)}(s)+Y^1_n A_{tot}^{1(n)}\right ) R_{\pi_2}(s_{\pi
N},q^2) \left(\varphi_n^+(\vec\sigma \vec p_\perp)\varphi_p \right)
g^{(\pi_2)}_{pn},\nn \\
\ee
where
\bea
A_{tot}^{0(n)}(s)&=& \frac{1}{s^2_{\pi N}}(k^{\perp q}_{3z})^2\sum\limits_J
\bigg [ R^0_n(P_2P''_J) C^{(J)}_3 A^{(J)}_{-2}(s) \bigg. \nn \\
&-&\frac 23 R^0_n(z(P'_J+2zP''_J)) C^{(J)}_2 A^{(J)}_{0}(s)\bigg ],
\nn \\
A_{tot}^{1(n)}(s)&=& \frac{1}{s^2_{\pi N}}
k^{\perp q}_{3z}k_{3x}\sum\limits_J \bigg [ 3R^1_n(P_2P''_J) C^{(J)}_3
A^{(J)}_{-2}(s) \bigg. \nn \\ &-&R^1_n(z(P'_J+2zP''_J)) C^{(J)}_2
A^{(J)}_{0}(s)\bigg ].
\eea
 Then the final expression is:
\bea
&&N(M,t){\langle{Y}^0_J\rangle}=\frac{\rho(s)\sqrt s}{\pi|\vec
p_2|^2s_{\pi N}} R_{\pi_2}(s_{\pi N},q^2) R^*_{\pi_2}(s_{\pi
N},q^2)(-t) (g^{(\pi_2)}_{pn})^2 \nn\\
&&\times\sum\limits_{n,m}\left [
d^{0\,0\,0}_{n,m,J}A_{tot}^{0(n)}(s)A_{tot}^{0(m)*}(s)+
d^{1\,1\,0}_{n,m,J}A_{tot}^{1(n)}(s)A_{tot}^{1(m)*}(s)\right ],
\nn \\
&&N(M,t){\langle{Y}^1_J\rangle}=\frac{\rho(s)\sqrt s}{\pi|\vec
p_2|^2s_{\pi N}} R_{\pi_2}(s_{\pi N},q^2) R^*_{\pi_2}(s_{\pi
N},q^2)(-t) (g^{(\pi_2)}_{pn})^2 \nn \\
&&\times\sum\limits_{n,m}\left [
d^{1\,0\,1}_{n,m,J}A_{tot}^{1(n)}(s)A_{tot}^{0(m)*}(s)+
d^{0\,1\,1}_{n,m,J}A_{tot}^{0(n)}(s)A_{tot}^{1(m)*}(s)\right ].
\eea

\section{Status of trajectories on $(J,M^2)$ plane}

\begin{figure}[ht]
\centerline{\epsfig{file=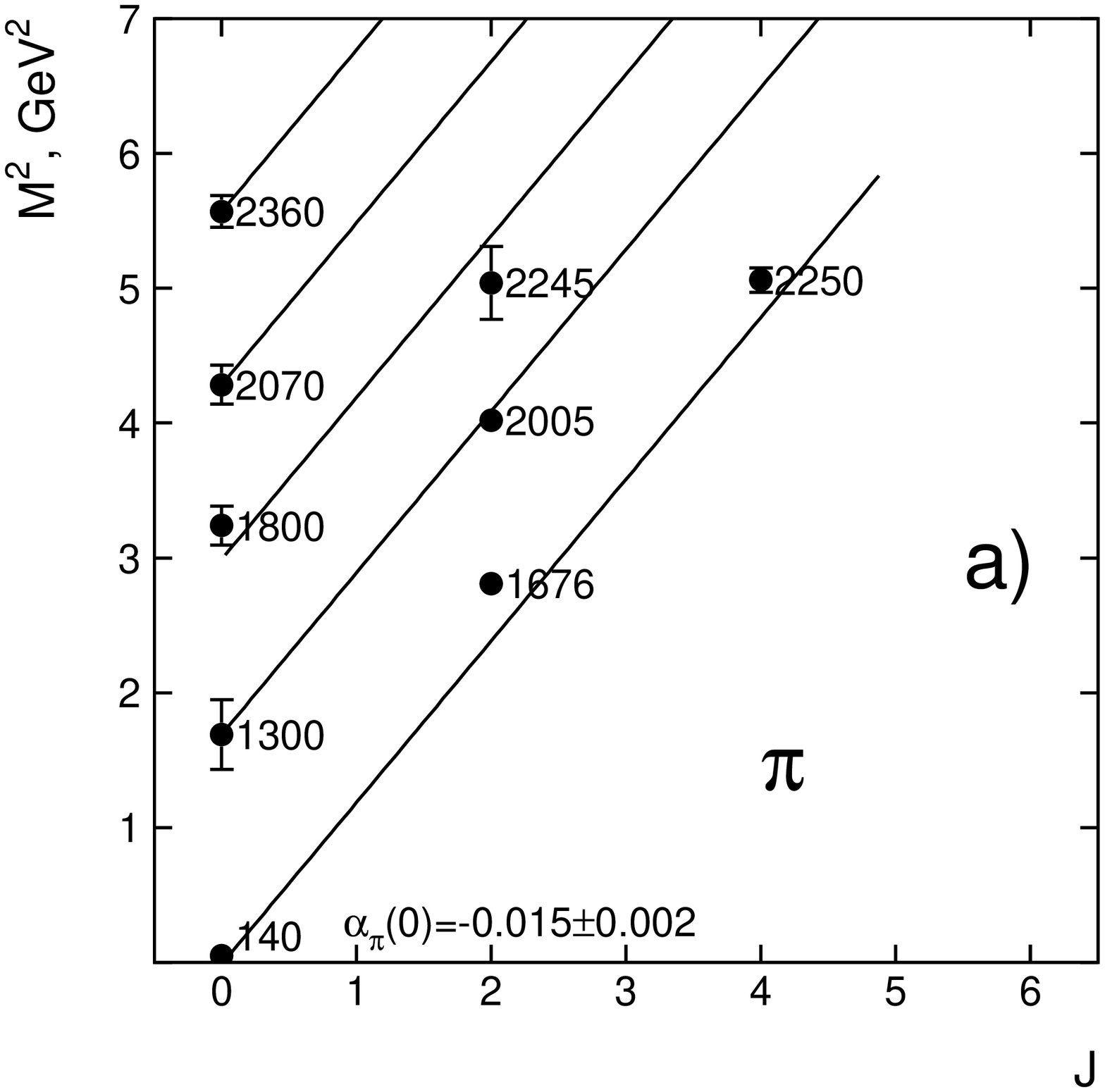,width=6.0cm}\hspace{-6mm}
            \epsfig{file=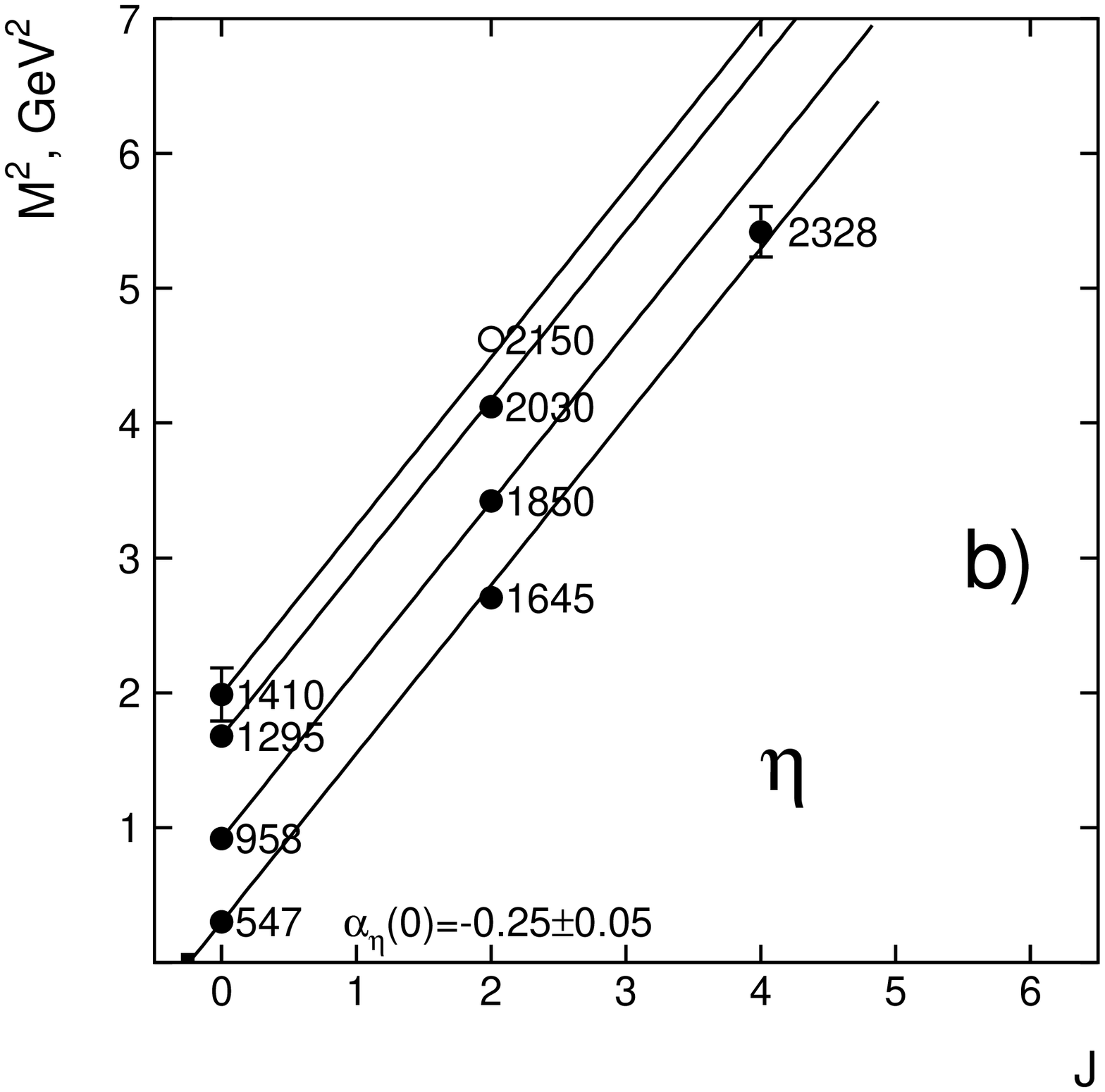,width=6.0cm}}
\vspace{-7mm}
\centerline{\epsfig{file=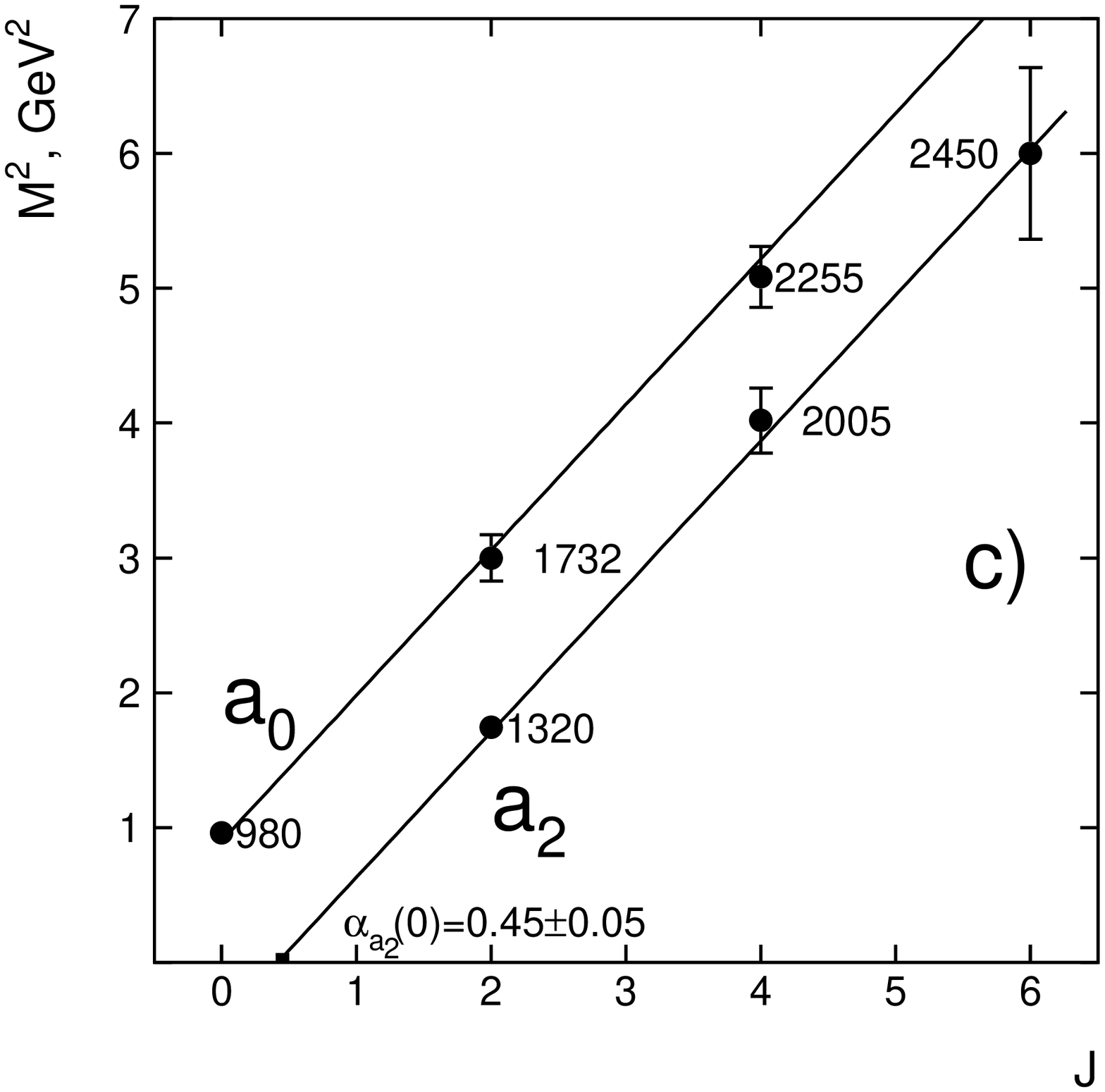,width=6.0cm}\hspace{-6mm}
            \epsfig{file=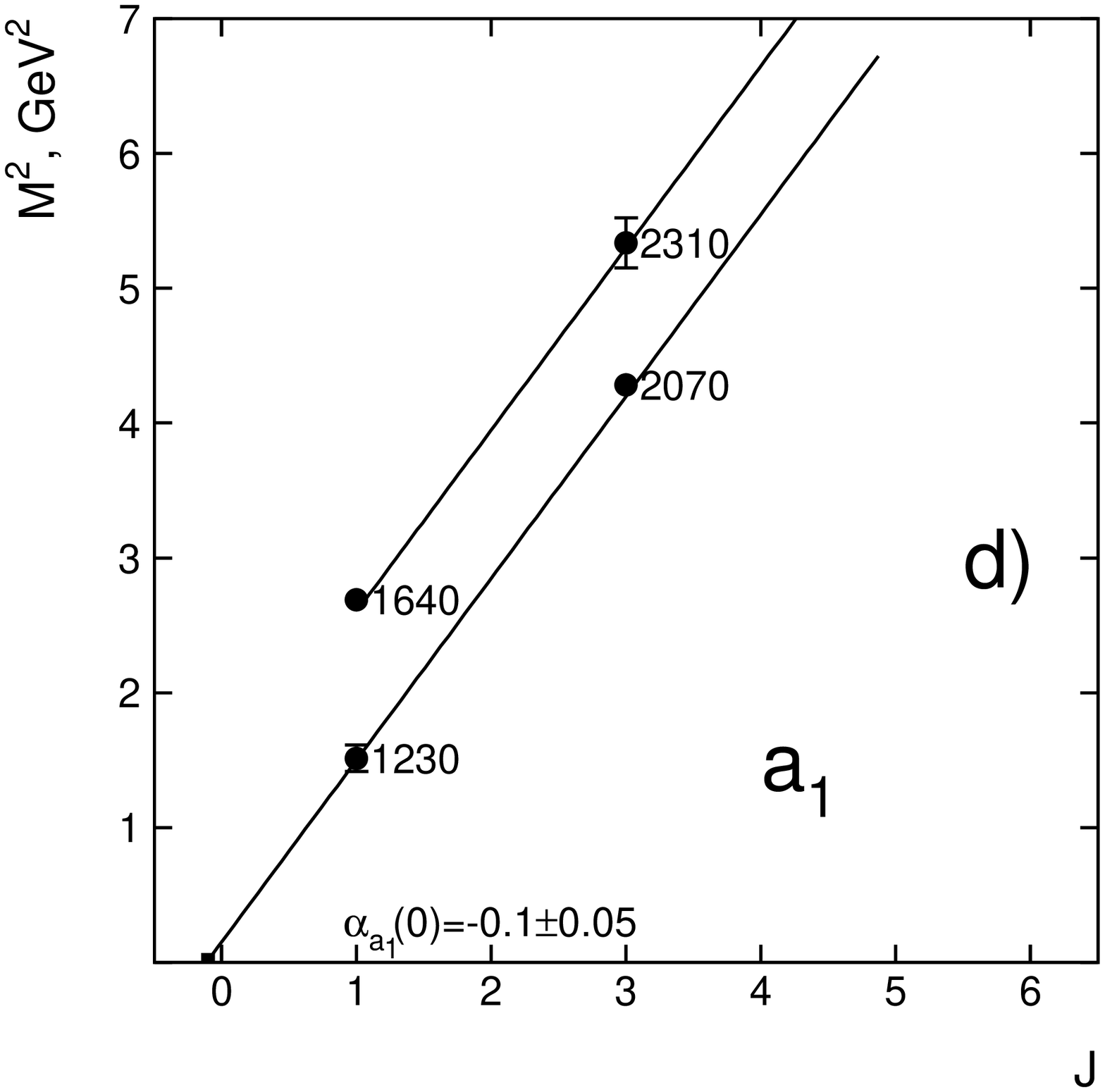,width=6.0cm}}
\vspace{-7mm}
\centerline{\epsfig{file=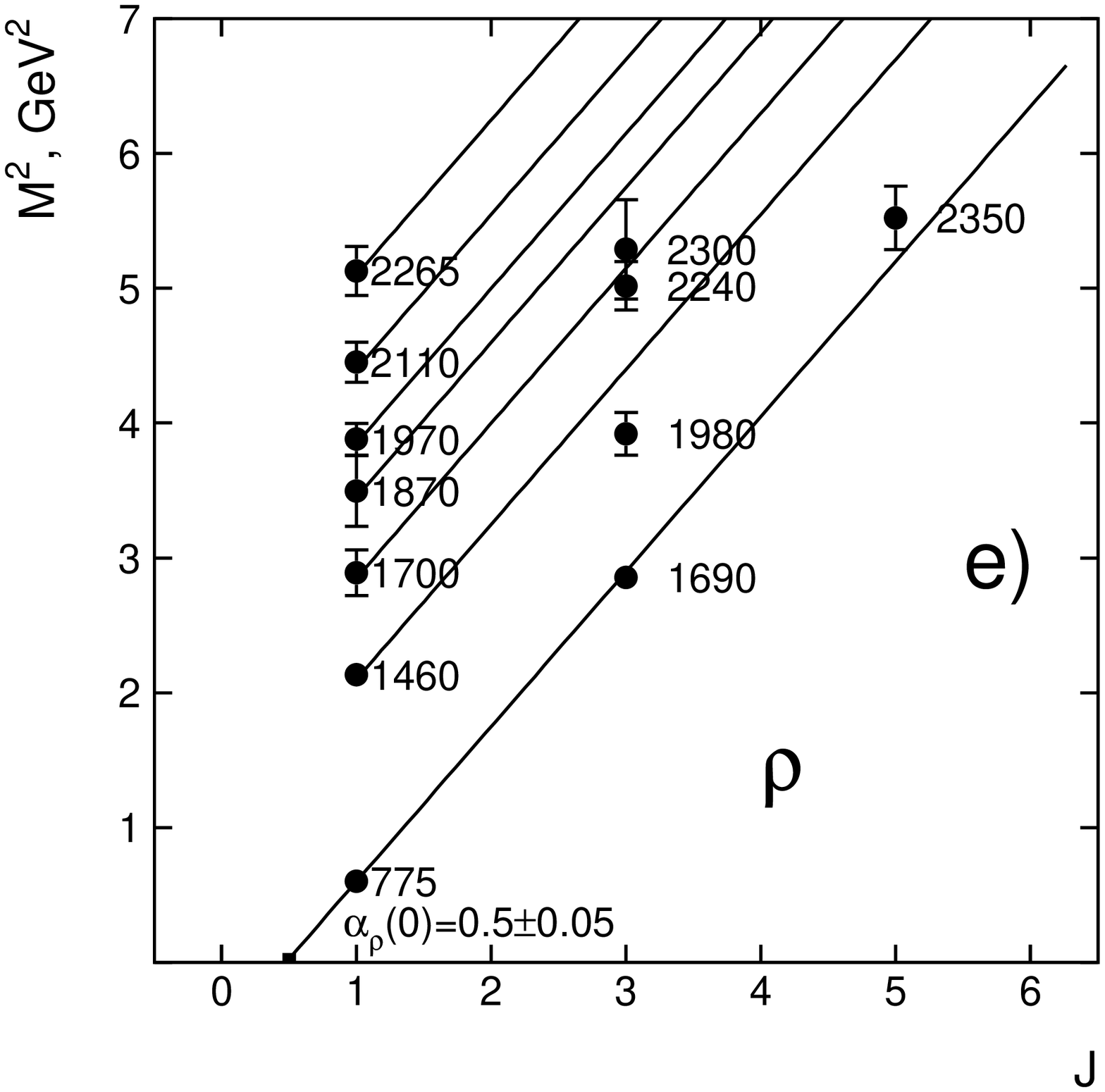,width=6.0cm}\hspace{-6mm}
           \epsfig{file=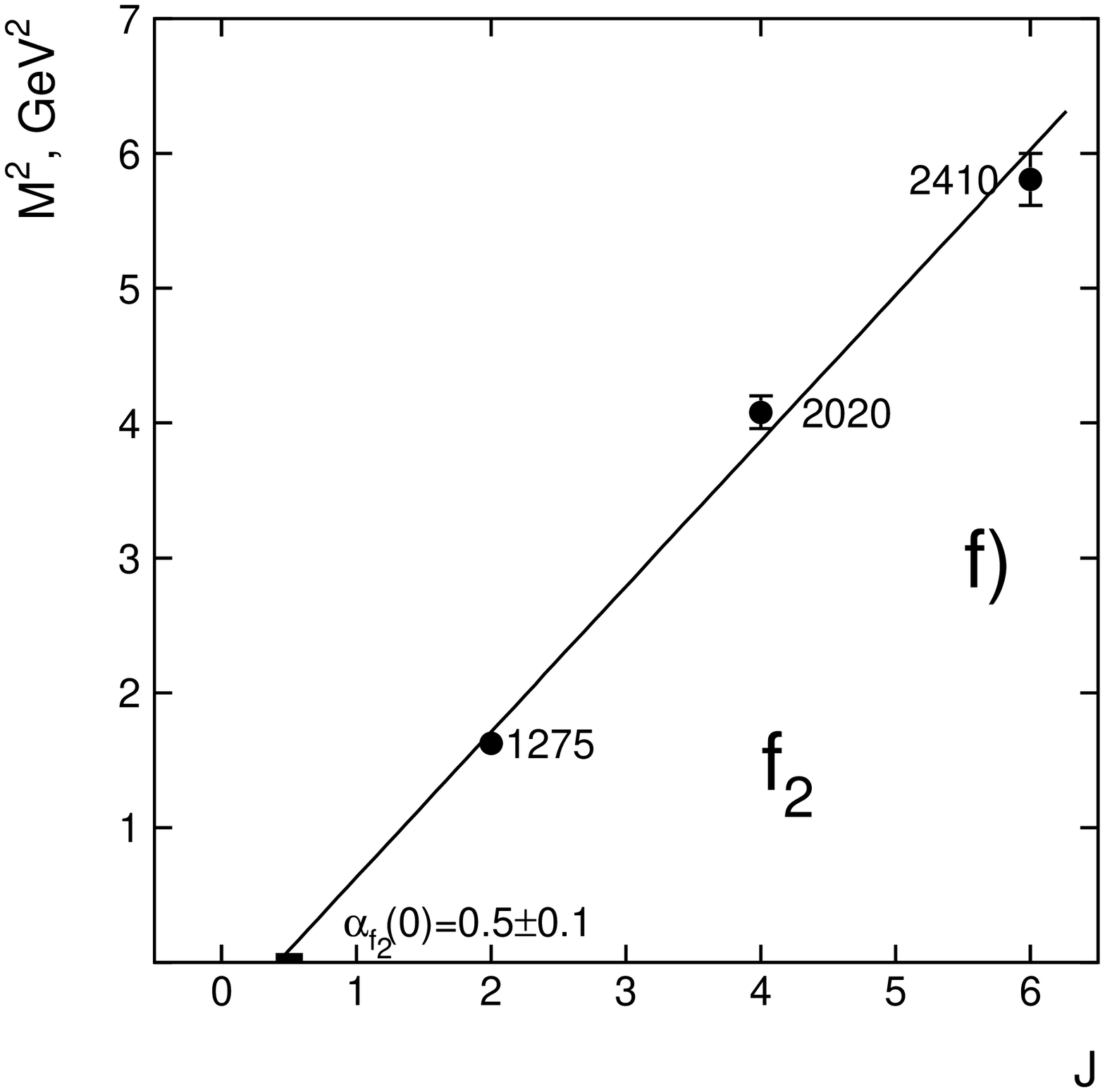,width=6.0cm}}
\vspace{-6mm}
\caption{Trajectories in the $(J, M^2)$ plane:
a) leading and daughter $\pi$-trajectories, b) leading and daughter $\eta$-trajectories,
c) $a_2$-trajectories, d) leading and daughter $a_1$-trajectories,
e) $\rho$-trajectories,
   f)~$P'$-trajectories.\label{X2f4} }
\end{figure}

 The $\pi$, $\eta$, $a_1$, $a_2$,  $a_3$, $\rho$ and $P'$ (or $f_2$)
 trajectories on $(J,M^2)$ planes are shown in Fig. \ref{X2f4}.

Leading $\pi$ and $\eta$ trajectories are unambiguously determined
together with their daughter trajectories, while for $a_2$, $a_1$,
$\rho$ and $P'$  only the leading trajectories can be given in a
definite way.

In the construction of $(J,M^2)$-trajectories it is essential
that the leading meson trajectories $(\pi, \rho, a_1, a_2$ and  $P')$
are well known from the analysis of the diffraction scattering of hadrons
at $p_{lab}\sim5-50$~GeV/c (for example, see \cite{book2} and
references therein).

The pion and $\eta$  trajectories are linear with a good accuracy (see
Fig.~\ref{X2f4}). Other leading trajectories ($\rho$, $a_1$, $a_2$,
$P'$) can also be considered as linear:
\begin{equation}
\label{X2e13}
\alpha_X(M^2)\ \simeq\ \alpha_X(0)+\alpha'_X(0)M^2\ .
\end{equation}
The parameters of the linear trajectories, determined by the masses of
the $q\bar{q}$ states, are
\begin{eqnarray}
\label{X2e14}
& \alpha_\pi(0)\simeq-0.015\ , & \alpha'_\pi(0)\simeq0.83 \;
\rm{GeV}^{-2}; \nonumber\\
& \alpha_\rho(0)\simeq0.50\ , & \alpha'_\rho(0)\simeq0.87 \;
\rm{GeV}^{-2}; \nonumber\\
& \alpha_\eta(0)\simeq-0.25\ , & \alpha'_\eta(0)\simeq0.80 \;
\rm{GeV}^{-2}; \nonumber\\
& \alpha_{a_1}(0)\simeq -0.10\ , & \alpha'_{a_1}(0)\simeq0.72 \;
\rm{GeV}^{-2}; \nonumber\\
& \alpha_{a_2}(0)\simeq0.45\ , & \alpha'_{a_2}(0)\simeq0.93 \;
\rm{GeV}^{-2}; \nonumber\\
& \alpha_{P'}(0)\simeq0.50\ , & \alpha'_{P'}(0)\simeq0.93 \;
\rm{GeV}^{-2}.
\end{eqnarray}
The slopes $\alpha'_X(0)$ of the trajectories are approximately
equal. The inverse slope, $1/\alpha'_X(0)\simeq 1.25\pm
0.15$~GeV$^2$, roughly equals the parameter $\mu^2$ for trajectories
on the $(n,M^2)$ planes:
\begin{equation}
\label{X2e15}
\frac1{\alpha'_X(0)}\ \simeq\ \mu^2\ .
\end{equation}
In the subsequent chapters, considering the scattering processes, we
use for the Regge trajectories the momentum transfer squared $M^2\to
t$.

\subsection{Kaon trajectories on $(J,M^2)$ plane}

\begin{figure}[ht]
\centerline{\epsfig{file=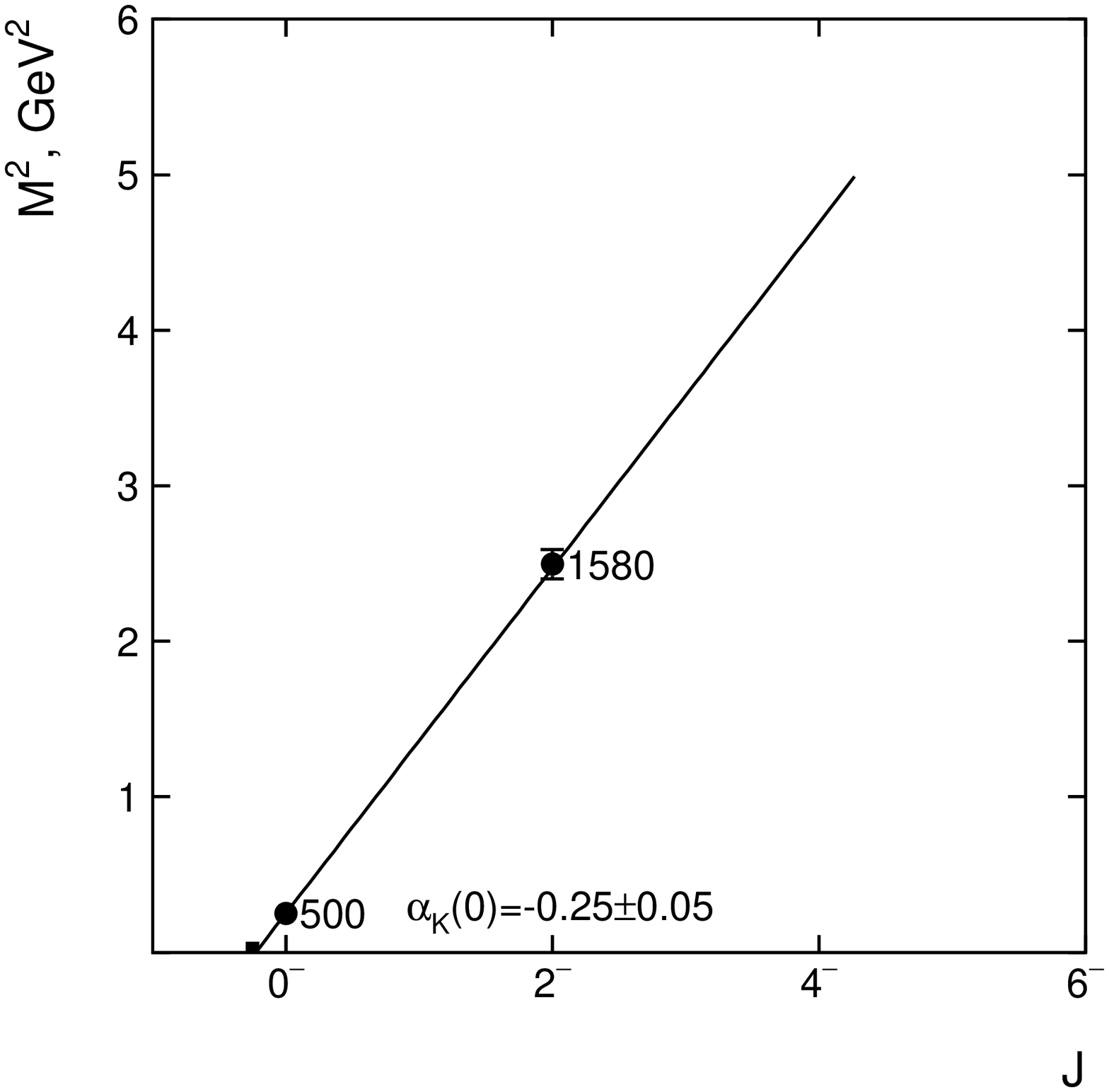,width=5.5cm}\hspace{-1.cm}
            \epsfig{file=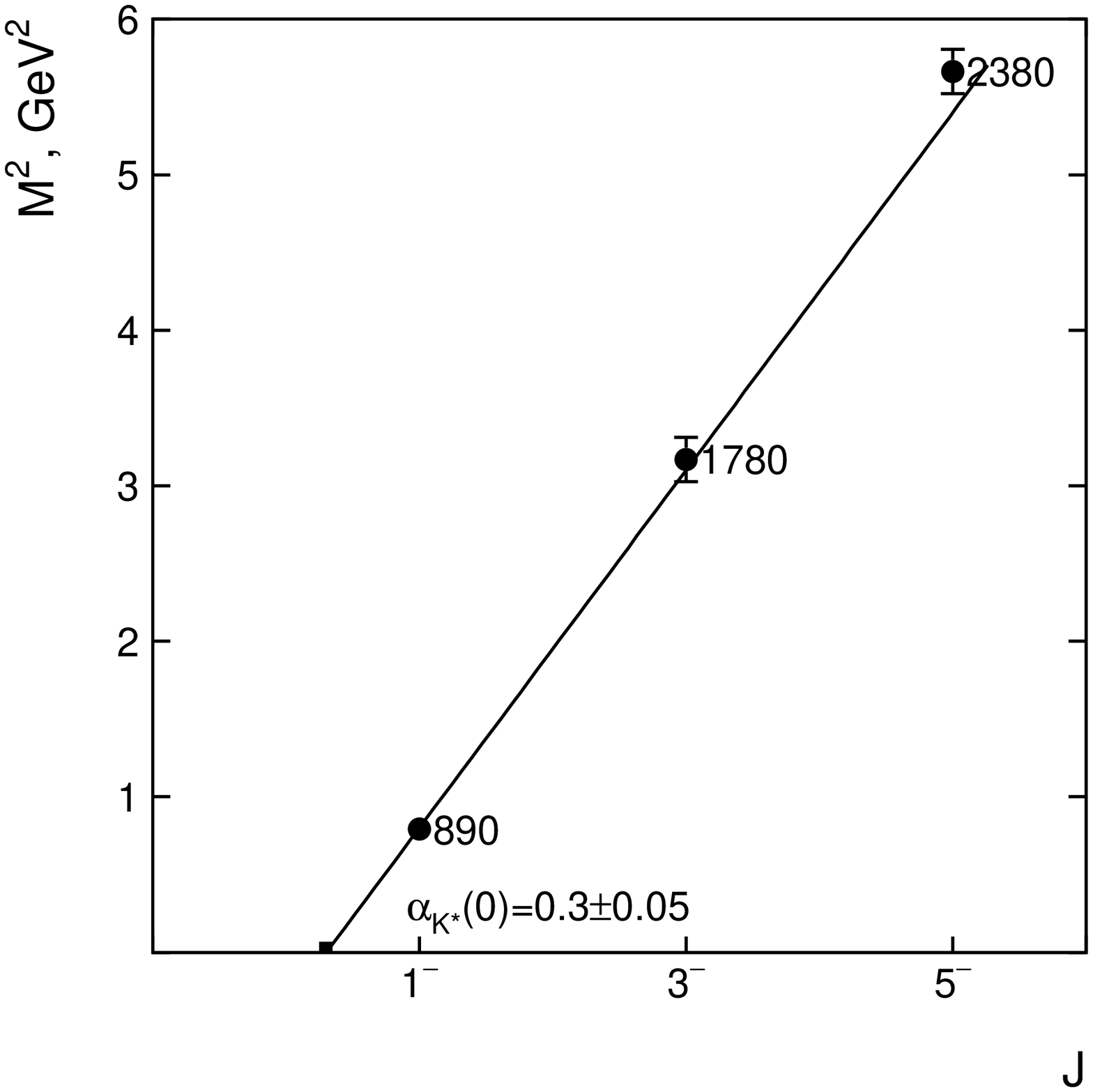,width=5.5cm}}
\vspace{-0.5cm}
\centerline{\epsfig{file=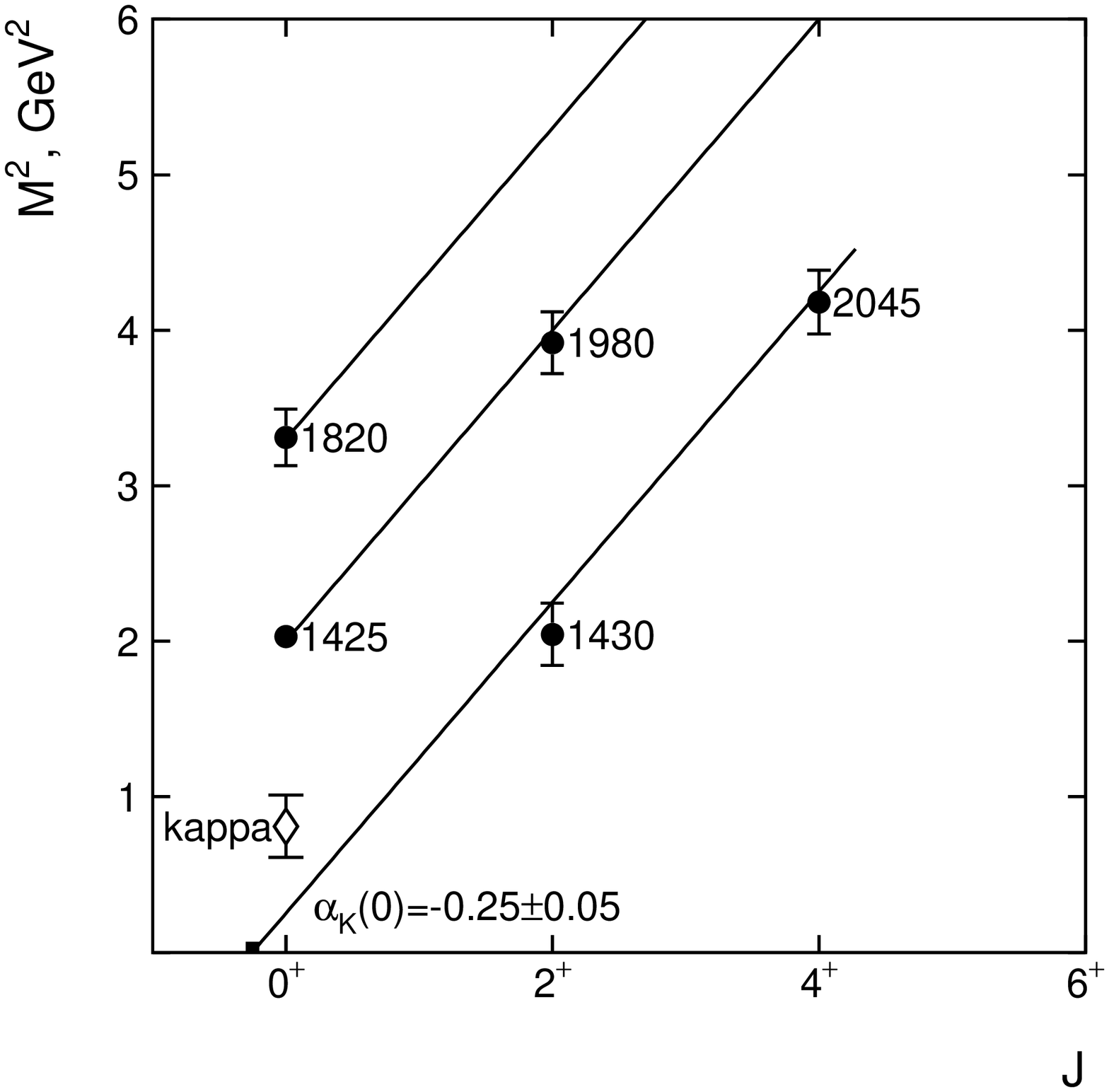,width=5.5cm}}
\vspace{-0.5cm}
\caption{Kaon trajectories on the $(J^P, M^2)$ plane.\label{X2f5}}
\end{figure}

As was said above, experimental data in the kaon sector are scarce,
 so in Fig. \ref{X2f5}
 we show only the leading $K$-meson trajectory
 (the states with $J^P=0^-$, $2^-$), the $K^*$ trajectory ($J^P=1^-,3^-,5^-$)
 and the leading and daughter trajectories for $J^P=0^+$, $2^+$, $4^+$.
The parameters of the leading kaon  trajectories are as follows:
\begin{eqnarray}
\label{X2e16}
& \alpha_K(0)\simeq-0.25\ , & \alpha'_K(0)\simeq 0.90 \;
\rm{GeV}^{-2}; \nonumber\\
& \alpha_{K^*}(0)\simeq 0.30\ , & \alpha'_{K^*}(0)\simeq 0.85 \;
\rm{GeV}^{-2}; \nonumber\\
& \alpha_{K_{2^+}}(0)\simeq -0.2\ , & \alpha'_{K_{2^+}}(0)\simeq 1.0 \;
\rm{GeV}^{-2}.
\end{eqnarray}

\clearpage

\clearpage

 The trajectories with  $J^P=1^+$, $3^+$, $5^+$ cannot be defined unambiguously.

\section{Dispersion integral equation for a three-body system
and the $K$-matrix approach}

The LEAR (CERN) experiment accumulated high statistics data on
three-meson production from the $\bar p p$ annihilation at rest,
mainly from ($J^{PC}=0^{-+}$)-level. The data of the Crystal Barrel
Collaboration (LEAR) were successfully analysed  in the terms of the
$K$-matrix approach (see, for example,
\cite{Tkm,Tf1500,Tcbc,TCB-94}) with the aim to search for new meson
resonances in the region 1000--1600 MeV. But the $K$-matrix approach
takes into account three-body unitarity condition on the
phenomenological level only. The three-body unitarity can be taken
into account considering all three-particle rescatterings, using,
for example, the dispersion relation $N/D$-method.

In this Appendix we demonstrate what type of assumptions is needed
for transformation of the three-body dispersion relation amplitudes
into that of the $K$-matrix approach.

Here the dispersion relation $N/D$-method is presented
for a three-body system: the method allows one to take into account
final-state two-meson interactions. We consider in detail an
illustrative example: the decay of the $0^{-+}$-state into three
different pseudoscalar mesons.

The first steps in  accounting for all two-body final state
interactions were made in \cite{TSTMD} in a non-relativistic approach
for three-nucleon systems.  In \cite{Tfaddeev} the two-body
interactions were considered in the potential approach (the Faddeev
equation).

The relativistic dispersion relation technique was used for the
investigation of the final state interaction effects in \cite{TAD}.

A relativistic dispersion relation equation for the amplitude
$\eta\to \pi\pi\pi$ was written in \cite{TAA-eta}. Later on the method
was generalised \cite{TAA-3mesons} for the coupled processes
 $p\bar p({\rm at\, rest})\to\pi\pi\pi$, $\eta\eta\pi$, $K\bar K\pi$:
this way a system of coupled equations for decay amplitudes was
written. Following \cite{TAA-eta,TAA-3mesons}, we explain here the
main points in considering the dispersion relations for a
three-particle system. The account of the three-particle final state
interactions imposes correct unitarity and analyticity constraints
on the amplitude.

\subsubsection{Two-particle interactions in the
 $0^-$-$state\longrightarrow  P_1P_2P_3$ decay}

\begin{figure}
\centerline{\epsfig{file=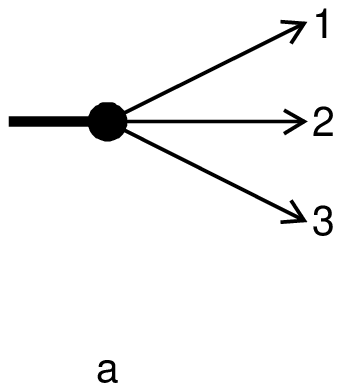,height=3cm}
            \epsfig{file=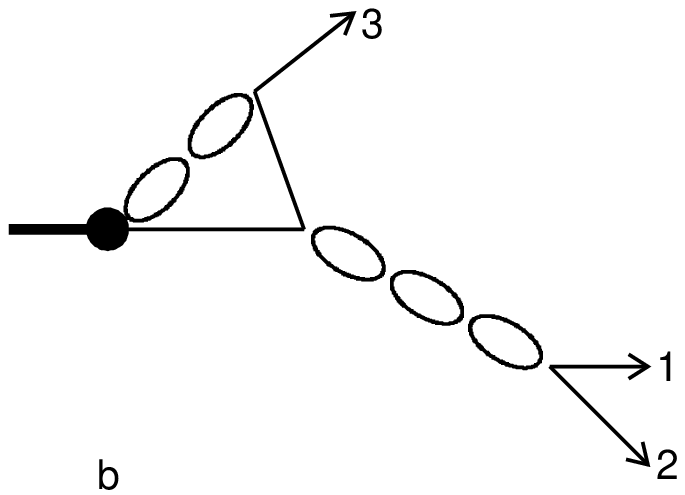,height=3cm}}
\caption{ Different types of transitions
 $(J^P_{in}=0^-)$-state$\longrightarrow  P_1P_2P_3$: a) prompt decay, b)
 decay with subsequent final state interactions.}
\label{Tcha6-5}
\end{figure}

As previously, we consider the decay of a pseudoscalar particle
$(J^P_{in}=0^-)$ with the mass $M$ and momentum $P$ into three
pseudoscalar particles with masses $m_1$, $m_2$, $m_3$ and momenta
$k_1$, $k_2$, $k_3$. There are different contributions
to this decay process: those without final state particle interactions
(prompt decay, Fig. \ref{Tcha6-5}a) and decays with subsequent final
state interactions (an example is shown in Fig. \ref{Tcha6-5}b).

For the decay amplitude we consider here an equation which takes
into account  two-particle final state interactions, such as that
shown in Fig. \ref{Tcha6-5}b. First, we consider in detail the
$S$-wave  interactions. This case clarifies the main points of the
dispersion relation approach for the three-particle interaction
amplitude. Then we discuss a scheme for generalising the equations
for the case of higher waves.

\begin{figure}
\centerline{\epsfig{file=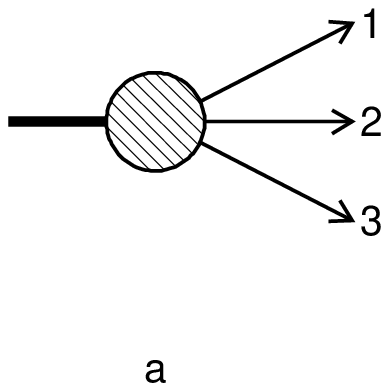,width=2.5cm}
            \epsfig{file=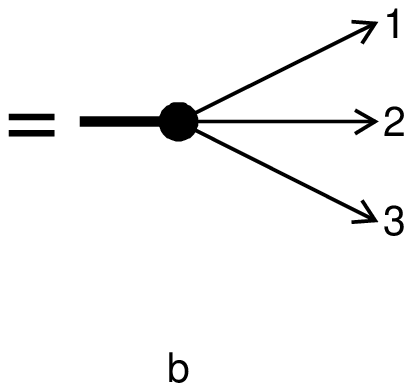,width=2.5cm}}
\centerline{\epsfig{file=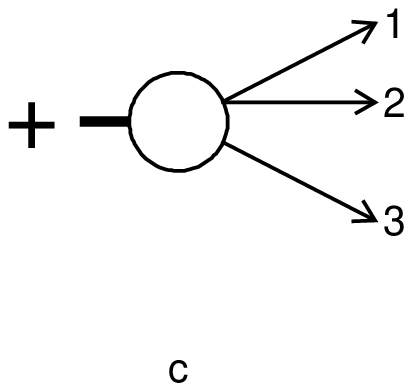,width=2.5cm}
            \epsfig{file=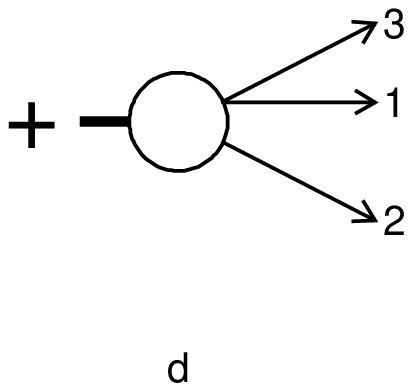,width=2.5cm}
            \epsfig{file=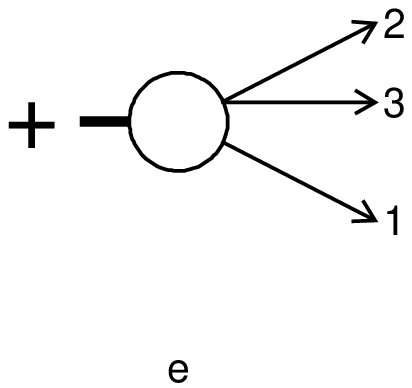,width=2.5cm}}
\caption{Different terms in the amplitude
 $A^{(J_{in}=0)}_{P_1P_2P_3}(s_{12},s_{13},s_{23})$} \label{Tcha6-6}
\end{figure}

{\bf (i) $S$-wave interaction.}

Let us begin with  the $S$-wave two-particle interactions.
The decay amplitude is given by
\be
A^{(J_{in}=0)}_{P_1P_2P_3}(s_{12},s_{13},s_{23})=
\lam(s_{12},s_{13},s_{23}) + A^{(0)}_{12}(s_{12}) +
A^{(0)}_{13}(s_{13}) +A^{(0)}_{23}(s_{23}).
\label{Tu4}
\ee
Different terms in (\ref{Tu4}) are illustrated by Fig. \ref{Tcha6-6}:
 we have a prompt production amplitude, Fig. \ref{Tcha6-6}b, and
terms $A^{(0)}_{ij}(s_{ij})$ with particles $P_1P_2$ (Fig.
\ref{Tcha6-6}c), $P_1P_3$ (Fig. \ref{Tcha6-6}d) and $P_2P_3$ (Fig.
\ref{Tcha6-6}e) participating in final state interactions.

To take into account rescattering of the type shown in Fig.
\ref{Tcha6-5}b, we can write equations for different terms
$A^{(0)}_{ij}(s_{ij})$.

The two-particle unitarity condition is explored to derive the
integral equation  for the amplitude $A^{(0)}_{ij}(s_{ij})$. The
idea of the approach suggested in \cite{TAA} is that one should
consider the case of a small external mass $M< m_1+m_2+m_3$. A
standard spectral integral equation (or a dispersion relation
equation) is written in this case for the transitions
 $h_{in}P_\ell\to P_iP_j$. Then the analytical continuation is
performed over the mass $M$ back to the decay region: this gives a
system of equations for decay amplitudes $A^{(0)}_{ij}(s_{ij})$.

So, let us consider the channel of particles 1 and 2, the transition
 $h_{in}P_3\to P_1P_2$.
We write the two-particle unitarity condition for the scattering in
this channel with the assumption $(M+m_3) \sim (m_1+m_2)$.

The discontinuity of the amplitude in the $s_{12}$-channel equals
\bea
\label{Tu5}
&&disc_{12}\, A^{J_{in}=0}_{P_1P_2P_3}
(s_{12},s_{13},s_{23}) = disc_{12}\, A^{(0)}_{12} (s_{12})=
 \\
&&=\!\int\!
d\Phi_{12}(p_{12};k_1,k_2) \biggl( \lam(s_{12},s_{13},s_{23}) +
A^{(0)}_{12}(s_{12}) + A^{(0)}_{13}(s_{13}) +A^{(0)}_{23}(s_{23})
\biggr)  \biggl(A_{12 \rightarrow
12}^{(0)}(s_{12})\biggr)^*   \ .
\nn
\eea
Here $d\Phi_{12}(p_{12};k_1,k_2)=(1/2) (2\pi)^{-2}
\delta^4(p_{12}-k_1-k_2)d^4 k_1 d^4 k_2 \delta(m_1^2-k_1^2)
\delta(m_2^2-k_2^2)$ is the standard phase volume of particles 1 and 2.
In (\ref{Tu5}), we should take into account that only
 $A^{(0)}_{12} (s_{12})$ has a non-zero discontinuity in the channel
$12$.

 {\it $S$-wave two-particle scattering amplitude }

But first, let us consider the $S$-wave two-particle scattering amplitude
$A^{(0)}_{P_1P_2\rightarrow P_1P_2}$. It  can be written
in the dispersion $N/D$ approach with separable interaction (see
Chapter 3) as a series
\begin{eqnarray} A^{(0)}_{P_1P_2\rightarrow P_1P_2}(s)
&=&
G_0^L(s_{12}) G_0^R(s_{12}) + G_0^L(s_{12}) B^{(0)}_{12}(s_{12}) G_0^R(s_{12})
 \\
&+& G_0^L(s_{12}) B^{(0)2}_{12}(s_{12}) G_0^R(s_{12})+ ... =
\frac{G_0^L(s_{12})
G_0^R(s_{12})}{1-B^{(0)}_{12}(s_{12})},~~~~~ \nn
\label{Tu6}
\end{eqnarray}
 where $G_0^L(s_{12})$ and
$G_0^R(s_{12})$ are left and right vertex functions. The loop
diagram $B^{(0)}_{12}(s_{12})$  in the dispersion relation
representation reads:
\be
\label{Tu7}
B^{(0)}_{12}(s_{12})=
\int\limits_{(m_1+m_2)^2}^{\infty} \frac{ds_{12}'}{\pi}
\frac{G_0^L(s_{12}') \rho^{(0)}_{12}(s_{12}')
G_0^R(s_{12}')}{s_{12}'-s_{12}-i0},
\ee
where
$\rho^{(0)}_{12}(s_{12})=
\sqrt{[s_{12}-(m_1+m_2)^2][s_{12}-(m_1-m_2)^2]}/(16\pi s_{12})$ is
the two-particle $S$-wave phase space integrated over the angular
 variables. The vertex functions contain left-hand singularities related
to the t-channel exchange diagrams, while the loop diagram
$B^{(0)}_{12}(s_{12})$ has a singularity due to the elastic
scattering (the right-hand side singularity). The consideration of the
scattering amplitude $A^{(0)}_{P_1P_2\rightarrow P_1P_2}(s_{12})$
does not specify it whether both vertices, $G_0^L(s_{12})$ and
$G_0^R(s_{12})$, have left-hand singularities or only one of them (see
discussion in Chapter 3). Considering the three-body decay, it is
convenient to make use of this freedom. On the first sheet of the
decay amplitude, we take into account the threshold  singularities
at $s_{ij}= (m_i+m_j)^2$, which are associated with the elastic
scattering in the subchannel of particles $i$ and $j$ but not those
on the left-hand side. This means that the vertex $G_0^R(s_{12})$
should be chosen here as an analytical function. For the sake of
simplicity let us put $G_0^R(s_{12})=1$ and present the amplitude
$P_1P_2\to P_1P_2$ as
 \be  \label{Tu8}
\hspace{-7mm}A^{(0)}_{P_1P_2\to P_1P_2}(s_{12})=
G_0^L(s_{12})\frac{1}{1-B^{(0)}_{12}(s_{12})}
\qquad
{\rm at} \quad
G_0^R(s_{12})=1\ .
 \ee

 {\it
 Equation for the
decay amplitude $h_{in}\to P_1P_2P_3$ }

Exploring (\ref{Tu5}), let us now return to the equation for the
decay amplitude $h_{in}\to P_1P_2P_3$.

As was noted (see also \cite{TAA}), the full set of rescattering of
particles 1 and 2 gives us the factor $(1-B_0(s_{12}))^{-1}$, so we
have from (\ref{Tu5}):
\be
A^{(0)}_{12}(s_{12})=B_{in}^{(0)}(s_{12})\frac{1}{1-B_0(s_{12})}\; .
 \label{Tu9}
\ee
The first loop diagram $B_{in}^{(0)}(s_{12})$ is determined as
\be
\label{Tu10}
 B_{in}^{(0)}(s_{12})= \int\limits_{(m_1+m_2)^2}^{\infty}
 \frac{ds'_{12}}{\pi}
\frac{disc_{12} \; B_{in}^{(0)}(s'_{12})}{s'_{12}-s_{12}-i0}\; ,
\ee
where
\begin{eqnarray}
\label{Tu11}
disc_{12} \; B_{in}^{(0)}(s_{12})
&=&\int
d\Phi_{12}(p_{12};k_1,k_2) \biggl( \lam(s_{12},s_{13},s_{23}) +
 A^{(0)}_{13}(s_{13}) +  A^{(0)}_{23}(s_{23})
 \biggr) G_0^L(s_{12}) \nonumber \\
&\equiv &disc_{12} \;  B_{\lambda -12}^{(0)}(s_{12})+
disc_{12} \; B_{13 -12}^{(0)}(s_{12}) +
disc_{12} \; B_{23 -12}^{(0)}(s_{12}).
\end{eqnarray}
Here we present $disc_{12} \;B_{in}^{(0)}(s_{12})$ as a sum of
three terms because each of them needs a special treatment when
$M^2+i\varepsilon$ is increasing.

 It is convenient to perform the phase-space integration in equation
(\ref{Tu11}) in the centre-of-mass system of particles 1 and 2 where
$\vk_1+\vk_2=0$. In this frame
\begin{eqnarray} \label{Tu12}
 &&
s_{13}= m_1^2 +m_3^2+ 2k_{10}k_{30} -2z\mid \vk_1\mid \mid \vk_3\mid\;,
\nn \\
&&   s_{23}= m_2^2 +m_3^2+ 2k_{20}k_{30} + 2z\mid \vk_2\mid
   \mid \vk_3\mid\;,
\end{eqnarray}
 where $z=\cos \theta_{13}$ and
$  k_{10} = \frac{s_{12}+m_1^2-m_2^2}{2\sqrt{s_{12}}}$,
$  k_{20} = \frac{s_{12}+m_2^2-m_1^2}{2\sqrt{s_{12}}}$,
$ -  k_{30} = \frac{s_{12}+m_3^2-M^2}{2\sqrt{s_{12}}}$.
The minus sign in front of $k_{30}$ reflects the fact that
$P_3$ is an outgoing, not an incoming particle. As usually,
 $ \mid \vk_j\mid=\sqrt{k_{j\, 0}^2-m_j^2}$ for $j=1,2,3$.
In the calculation of $disc_{12} \; B^{(in)}_0(s_{12})$ all
integrations are carried out easily except for the contour
integral over $dz$. It can be rewritten in (\ref{Tu11}) as an
integral over $ds_{13}$ or $ds_{23}$:
\be \label{Tu15}
\hspace{-6mm}
\int\limits^{+1}_{-1}\frac{dz}{2}\rightarrow
\int\limits^{s_{13}(+)}_{s_{13}(-)} \frac{ds_{13}}{ 4\mid \vk_1\mid
\mid \vk_3\mid},\quad {\rm or}\quad
\int\limits^{+1}_{-1}\frac{dz}{2}\rightarrow
\int\limits^{s_{23}(+)}_{s_{23}(-)} \frac{ds_{23}}{ 4\mid \vk_2\mid
\mid \vk_3\mid}, \ee
where
\begin{eqnarray}
\label{Tu16}
&&   s_{13}(\pm)= m_1^2 +m_3^2+ 2k_{10}k_{30} \pm 2\mid \vk_1\mid
\mid \vk_3\mid\;,
\nn\\
&&   s_{23}(\pm)= m_2^2 +m_3^2+ 2k_{20}k_{30}\pm 2\mid \vk_2\mid
   \mid \vk_3\mid\ .
\end{eqnarray}
The relative location of the integration contours (\ref{Tu15}) and
amplitude singularities is the determining point for writing the
equation.

Below we use the notation
\be \label{Tu15a}
\int\limits^{s_{i3}(+)}_{s_{i3}(-)}
ds_{i3}=\int\limits_{C_{i3}(s_{12})}  ds_{i3}.
\ee
 One can see from (\ref{Tu16}) that the integration contours $C_{13}(s_{12})$ and
$C_{23}(s_{12})$ depend on $ M^2$ and $s_{12}$, so we should monitor
them when  $M^2+i\varepsilon$ increases.

Let us underline again that the idea to consider the decay processes
in the dispersion relation approach is the following : we write the
equation in the region of the standard scattering $two\; particles
\to two\; particles$ (when $m_1\sim m_2\sim m_3\sim M $) with the
subsequent analytical continuation (with $M^2+i\varepsilon$ at
$\varepsilon >0$) into the decay region, $ M> m_1+ m_2+ m_3$, and
then $\varepsilon \to +0$. In this continuation we need to specify
what type of singularities (and corresponding type of processes) we
take into account and what type of singularities we neglect.
Definitely, we take into account right-hand side and left-hand side
singularities of the scattering processes $P_i P_j\to P_iP_j$ (our
main aim is to restore the rescattering processes correctly). But
singularities of the prompt production amplitude are beyond the
field of our interest. In other words, we suppose
$\lam(s_{12},s_{13},s_{23})$ to be an analytical function in the
region under consideration.

Assuming $\lam(s_{12},s_{13},s_{23})$ to be an analytical function
in the region under consideration, we can easily perform analytical
continuation of the integral  over $dz$, Eq. (\ref{Tu15}), with
$M^2+i\varepsilon$.

Problems may appear in the integrations of $A^{(0)}_{13}(s_{13})$
and $A^{(0)}_{23}(s_{23})$ owing to the threshold singularities in
the amplitudes (at $s_{13}=(m_1+m_3)^2$ and $s_{23}=(m_2+m_3)^2$,
respectively). However, the analytical continuation over
$M^2+i\varepsilon$ resolves them: one can see in \cite{book3} (Chapter 4)
 the location of the integration contour in the complex-$s_{23}$
plane with respect to the threshold singularity at
$s_{23}=(m_2+m_3)^2$ when $ M> m_1+m_2+ m_3$.

Let us now write the equation for the three-particle production
amplitude in more detail. We denote the $S$-wave projection of $\lam
(s_{12},s_{13},s_{23})$ as
\be \label{Tu17}
\langle\lam(s_{12},s_{13},s_{23})\rangle^{(0)}_{12}=
\int\limits_{-1}^{+1} \frac{dz}{2}\;\lam(s_{12},s_{13},s_{23}),
\ee
and the contour integrals over the amplitudes $A^{(0)}_{13}(s_{13})$
and $A^{(0)}_{23}(s_{23})$ as
\be \hspace{-7mm}\langle
A^{(0)}_{i3}(s_{i3})\rangle^{(0)}_{12} =
 \int\limits_{-1}^{+1} \frac{dz}{2}\;A^{(0)}_{i3}(s_{i3})
\equiv\!\!\!
 \int\limits_{C_{i3}(s_{12})}\!\!\! \frac{ds_{i3}}{ 4|\vk_i||\vk_3|}
A^{(0)}_{i3}(s_{i3}),\quad
 i=1,2.
\label{Tu18}
\ee
 Remind once more that the definition of the contours $C_i(s_{12})$ is
given in (\ref{Tu15a}) while the relative position of the contour
$C_2(s_{12})$ and the threshold singularity in the $s_{23}$-channel
is shown in Fig. 4.26. So, we rewrite  (\ref{Tu11}) in the form
\be
disc_{12}\; B^{(0)}_{in}(s_{12})& =& \biggl(
\langle\lam(s_{12},s_{13},s_{23})\rangle^{(0)}_{12} + \langle
A^{(0)}_{13}(s_{13}) \rangle^{(0)}_{12} +
 \langle A^{(0)}_{23}(s_{23}) \rangle^{(0)}_{12}
 \biggr) \nn \\
&\times&\rho^{(0)}_{12}(s_{12}) G_0^L(s_{12}),
 \label{Tu19}
\ee
Equation (\ref{Tu19}) allows us to write the dispersion integral
for the loop amplitude $ B^{(0)}_{in}(s_{12})$. As a result, we
have:
 \be  \label{Tu20}
\hspace{-7mm}A^{(0)}_{12}(s_{12})\!=\!
\biggl ( B^{(0)}_{\lambda -12}(s_{12})\!+\!B^{(0)}_{13 -12}(s_{12})
 \!+\!B^{(0)}_{23 -12}(s_{12})
\biggl )
\frac{1}{1-B^{(0)}_{12}(s_{12})}
\ee
where
\begin{eqnarray} \label{Tu21b}
&& B^{(0)}_{\lambda -12}(s_{12})=
  \int\limits_{(m_1+m_2)^2}^{\infty}
 \frac{ds'_{12}}{\pi}
 \langle\lam(s'_{12},s'_{13},s'_{23})\rangle^{(0)}_{12}
\frac{\rho^{(0)}_{12}(s'_{12})} {s'_{12}- s_{12}-i0}
G_0^L(s'_{12}) ,\nonumber \\
&&
B^{(0)}_{i3 -12}(s_{12})=  \int\limits_{(m_1+m_2)^2}^{\infty}
 \frac{ds'_{12}}{\pi}
 \langle A^{(0)}_{i3}(s'_{i3}) \rangle^{(0)}_{12}
\frac{\rho^{(0)}_{12}(s'_{12})} {s'_{12}- s_{12}-i0}
G_0^L(s'_{12}).
\label{Tu20a}
\end{eqnarray}
Let us emphasise that in the integrand  (\ref{Tu20a}) the energy squared is
 $s'_{12}$ and hence, calculating
$ \langle\lam(s'_{12},s'_{13},s'_{23})\rangle^{(0)}_{12}$ and
$\langle A^{(0)}_{i3}(s'_{i3}) \rangle^{(0)}_{12}$, we should use
Eqs. (\ref{Tu12})  -- (\ref{Tu15a}) with the replacement $s_{12}\to
s'_{12}$.

\begin{figure}
\centerline{\epsfig{file=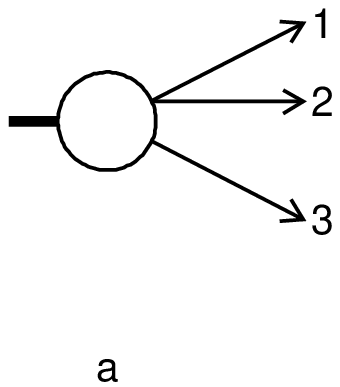,height=3cm}
            \epsfig{file=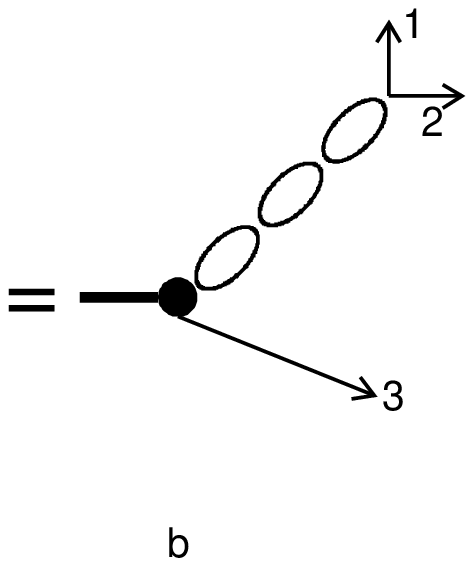,height=3cm}}
\centerline{\epsfig{file=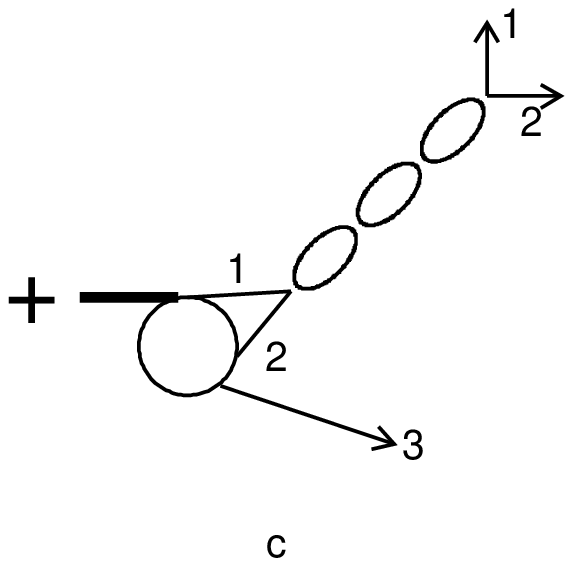,height=3cm}
            \epsfig{file=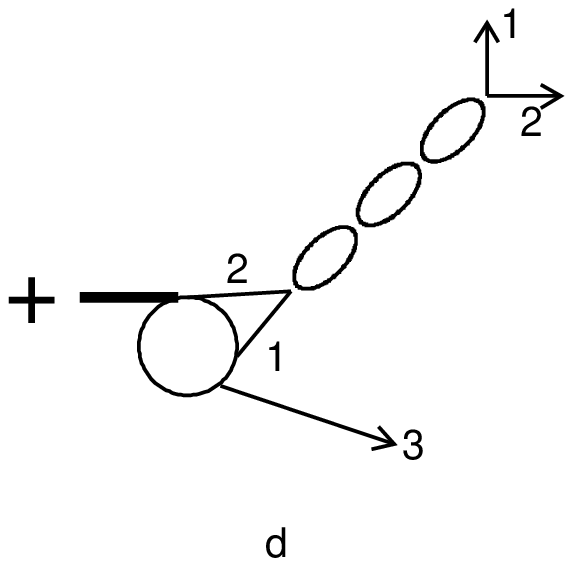,height=3cm}}
\caption{Diagrammatic presentation of Eq. (\ref{Tu20}).
}\label{Tcha6-8}
\end{figure}

The equation (\ref{Tu20}) is illustrated by Fig. \ref{Tcha6-8}.

In the same way we can write equations for $A^{(0)}_{13}(s_{13})$
and $A^{(0)}_{23}(s_{23})$. We have a system  of three
non-homogeneous equations which determine the amplitudes
$A^{(0)}_{ij}(s_{ij})$ when $\lam (s_{12},s_{13},s_{23})$ is
considered as an input function.

Note that the integration contour $C_i(s_{12})$ in (\ref{Tu18}), see
also \cite{TAA}, does not coincide with that of \cite{TAitchison66}
where the corresponding problem was treated starting from the
consideration of the three-body channel.

{\bf (ii) Final state rescattering $P_iP_j\to P_iP_j$ in the $L>0$
state.}

Equations for amplitudes which describe the final state interactions
in the transition $(J^P_{in}=0^-)$-state$\longrightarrow  P_1P_2P_3$
when rescattering $P_iP_j\to P_iP_j$ occur in a state with $L>0$
can be written in a way analogous to that presented above for $L=0$.
So, we suppose that $P_iP_j\to P_iP_j$ rescattering take place in a
state with definite orbital momentum $L$ and $L\neq 0$.

The amplitude for the decay $(J^P_{in}=0^-)$-state$\longrightarrow
P_1P_2P_3$ (below $L=J$) reads:
\begin{eqnarray}
&&A^{(J_{in}=0)}_{P_1P_2P_3}(s_{12},s_{13},s_{23})=
\lam(s_{12},s_{13},s_{23})
+ A^{(J)}_{12}(s_{12})
X^{(J)}_{\mu_1...\mu_{J}}(k_{3}^{\perp P})
 X^{(J)}_{\mu_1...\mu_{J}}(k_{12}^{\perp p_{12}})
\nn \\&&+ A^{(J)}_{13}(s_{13}) X^{(J)}_{\mu_1...\mu_{J}}(k_{2}^{\perp P})
X^{(J)}_{\mu_1...\mu_{J}}(k_{13}^{\perp p_{13}})
+A^{(J)}_{23}(s_{23}) X^{(J)}_{\mu_1...\mu_{J}}(k_{1}^{\perp P})
X^{(J)}_{\mu_1...\mu_{J}}(k_{23}^{\perp p_{23}})
.         \label{Tu21}
\end{eqnarray}
Convolutions of the momentum operators, such as
$X^{(J)}_{\mu_1...\mu_{J}}(k_{3}^{\perp P})$
$X^{(J)}_{\mu_1...\mu_{J}}(k_{12}^{\perp p_{12}})$,
 being functions of $s_{ij}$  do not contain
threshold singularities. So we can rewrite (\ref{Tu21}) in a more
compact form
\begin{eqnarray}
&&
A^{(J_{in}=0)}_{P_1P_2P_3}(s_{12},s_{13},s_{23})=
\lam(s_{12},s_{13},s_{23})
+ A^{(J-J)}_{12}(s_{12},s_{13},s_{23})
\nonumber \\
&&
+ A^{(J-J)}_{13}(s_{12},s_{13},s_{23})
+A^{(J-J)}_{23}(s_{12},s_{13},s_{23}),
         \label{Tu21a}
\end{eqnarray}
where
$ A^{(J-J)}_{12}(s_{12},s_{13},s_{23})=
 A^{(J)}_{12}(s_{12})
X^{(J)}_{\mu_1...\mu_{J}}(k_{3}^{\perp P})
 X^{(J)}_{\mu_1...\mu_{J}}(k_{12}^{\perp p_{12}})$,
and so on. As previously, $\lam(s_{12},s_{13},s_{23})$ is an
analytical function of $s_{ij}$ while the terms
$A^{(J-J)}_{ij}(s_{12},s_{13},s_{23})$ have threshold singularities
of the type $\sqrt{s_{ij}-(m_i+m_j)^2}$ due to final state
rescattering $P_iP_j\to P_iP_j$.

 {\it Two-particle scattering amplitude at $L=J\neq 0$}

First, we should introduce a two-particle $(L=J)$-wave scattering
amplitude. To be definite, we consider $P_1P_2\to P_1P_2$. We write
the block of a one-fold scattering as
\be
A^{(J)}_{(P_1P_2\to
P_1P_2)_{\rm one-fold}}(s_{12})&=&
X^{(J)}_{\nu_1...\nu_{J}}(k'^{\perp p_{12}}_{12})
 O^{\nu_1...\nu_{J}}_{\mu_1...\mu_{J}}(\perp p_{12})
G_J^L(s_{12}) G_J^R(s_{12})
\nn \\
&\times&X^{(J)}_{\mu_1...\mu_{J}}(k_{12}^{\perp p_{12}}).
\label{Tu22}
\ee
 The two-fold scattering amplitude reads:
\begin{eqnarray} \label{Tu22-2}
&&
A^{(J)}_{(P_1P_2\to P_1P_2)_{\rm two-fold}}(s_{12})=
X^{(J)}_{\nu_1...\nu_{J}}(k'^{\perp p_{12}}_{12})
 O^{\nu_1...\nu_{J}}_{\nu'_1...\nu'_{J}}(\perp p_{12})G_J^L(s_{12})
 \nn \\
&&\times
\biggl[\int\limits^\infty_{(m_1+m_2)^2}\frac{ds''_{12}}{\pi(s''_{12}-s_{12}-i0)}
G_J^R(s''_{12})\biggr.
\nn \\
&&\times\biggl.\int X^{(J)}_{\nu'_1 ...\nu'_J} (k''^{\perp p''_{12}}_{12})
 d\Phi_{12} (p''_{12};k''_1\, , k''_2)
X^{(J)}_{\nu''_1 ...\nu''_J} (k''^{\perp p''_{12}}_{12})G_J^L(s''_{12})
\biggr ]
\nonumber\\
&&\times
G_J^R(s_{12}) O^{\nu''_1...\nu''_{J}}_{\mu_1...\mu_{J}}(\perp
p_{12})G_J^L(s_{12}) X^{(J)}_{\mu_1...\mu_{J}}(k_{12}^{\perp p_{12}}).
 \end{eqnarray}
In the integrand we can replace $k''^{\perp p''_{12}}_{12}\to
k''^{\perp p_{12}}_{12}$, because in the c.m. frame of particles
$P_1P_2$ one has $p''_{12}=(\sqrt{s''_{12}},0,0,0)$ and
$p_{12}=(\sqrt{s_{12}},0,0,0)$. The integration over the phase space
gives \be \label{Tu23} \int X^{(J)}_{\nu'_1 ...\nu'_J} (k''^{\perp
p_{12}}_{12})
 d\Phi_{12} (p''_{12};k''_1\, , k''_2)
X^{(J)}_{\nu''_1 ...\nu''_J} (k''^{\perp p_{12}}_{12})&=&
O^{\nu'_1 ...\nu'_J}_{\nu''_1 ...\nu''_J} (\perp p_{12})
\nn \\&\times& \rho^{(J)}_{12} (s''_{12}).
\ee
 Using
\be \label{Tu24}
\hspace{-7mm} O^{\nu_1...\nu_{J}}_{\nu'_1...\nu'_{J}}(\perp p_{12})
O^{\nu'_1 ...\nu'_J}_{\nu''_1 ...\nu''_J} (\perp p_{12})
 O^{\nu''_1...\nu''_{J}}_{\mu_1...\mu_{J}}(\perp p_{12}) =
 O^{\nu_1...\nu_{J}}_{\mu_1...\mu_{J}}(\perp p_{12}),
\ee
we write the two-fold amplitude as follows:
\begin{eqnarray} \label{Tu25}
&&
A^{(J)}_{(P_1P_2\to P_1P_2)_{\rm two-fold}}(s_{12})= \\
&& X^{(J)}_{\nu_1...\nu_{J}}(k'^{\perp p_{12}}_{12})
 O^{\nu_1...\nu_{J}}_{\mu_1...\mu_{J}}(\perp p_{12})G_J^L(s_{12})
B^{(J)}_{12}(s_{12})
G_J^R(s_{12}) X^{(J)}_{\mu_1...\mu_{J}}(k_{12}^{\perp p_{12}}) ,
\nonumber
 \end{eqnarray}
where
\be \label{Tu26}
\hspace{-7mm}B^{(J)}_{12}(s_{12}) =
\int\limits^\infty_{(m_1+m_2)^2}\frac{ds''_{12}}{\pi(s''_{12}-s_{12}-i0)}
G_J^R(s''_{12})\rho^{(J)}_{12} (s''_{12})
G_J^L(s''_{12})
\ee
is the loop diagram.

The full set of rescattering gives:
\begin{eqnarray} \label{Tu27}
\hspace{-7mm}
A^{(J)}_{P_1P_2\to P_1P_2}(s_{12})&= &
X^{(J)}_{\nu_1...\nu_{J}}(k'^{\perp p_{12}}_{12})
 O^{\nu_1...\nu_{J}}_{\mu_1...\mu_{J}}(\perp p_{12})
\nn \\ &\times&
G_J^L(s_{12})
\frac{1}{1-B^{(J)}_{12}(s_{12})}
G_J^R(s_{12}) X^{(J)}_{\mu_1...\mu_{J}}(k_{12}^{\perp p_{12}}).
 \end{eqnarray}
As previously for $J=0$, we put
 \be  \label{Tu28}
  G_J^R(s_{12})=1\,   .
  \ee
Finally we write:
\be \label{Tu29}
A^{(J)}_{P_1P_2\to P_1P_2}(s_{12})=
 X^{(J)}_{\mu_1...\mu_{J}}(k'^{\perp p_{12}}_{12})
 G_J^L(s_{12})
\frac{1}{1-B^{(J)}_{12}(s_{12})}
 X^{(J)}_{\mu_1...\mu_{J}}(k_{12}^{\perp p_{12}}).\nn\\
 \ee

 {\it
 Equation for the
decay amplitude $h_{J_{in}=0}\to P_1P_2P_3$ }

Let us return now to the equation for the three-particle production
amplitude $A^{(J_{in}=0)}_{P_1P_2P_3}(s_{12},s_{13},s_{23})$ given
by (\ref{Tu21a}). We write equations for separated terms
$A^{(J-J)}_{ij}(s_{12},s_{13},s_{23})$. To use Eqs.
(\ref{Tu22})--(\ref{Tu29}) directly, we consider the term with the final state
interaction in the channel $12$, namely,
$A^{(J-J)}_{12}(s_{12},s_{13},s_{23})$.

The amplitude $A^{(J-J)}_{12}(s_{12},s_{13},s_{23})$ is determined
by three terms shown in Fig. \ref{Tcha6-8}b,c,d.

The term initiated by the prompt production block
$\lam(s_{12},s_{13},s_{23})$ is a set of loop diagrams of the type
of that in Fig. \ref{Tcha6-8}b. Therefore this term reads:
\be \label{Tu30}
 X^{(J)}_{\mu_1...\mu_{J}}(k^{\perp p_{12}}_{3})
B^{(J)}_{\lambda -12}(s_{12})\frac{1}{1-B^{(J)}_{12}(s_{12})}
 X^{(J)}_{\mu_1...\mu_{J}}(k_{12}^{\perp p_{12}}),
 \ee
with
\be
\hspace{-5mm}B^{(J)}_{\lambda -12}(s_{12})\!=\hspace{-3mm}
  \int\limits_{(m_1+m_2)^2}^{\infty}\hspace{-3mm}
 \frac{ds'_{12}}{\pi}
 \langle\lam(s'_{12},s'_{13},s'_{23})\rangle^{(J)}_{12}
\frac{\rho^{(J)}_{12}(s'_{12})} {s'_{12}- s_{12}-i0} G_0^L(s'_{12})
\label{Tu30a}.
\ee
Let us explain Eqs. (\ref{Tu30}),  (\ref{Tu30a}) in
more detail. Similarly to (\ref{Tu22}), we write for the first loop
diagram in (\ref{Tu30}) the following representation:

\begin{eqnarray} \label{Tu30b}
&&\hspace{-2mm}
 X^{(J)}_{\mu_1...\mu_{J}}(k^{\perp p_{12}}_{3})
B^{(J)}_{\lambda -12}(s_{12})
 X^{(J)}_{\mu_1...\mu_{J}}(k_{12}^{\perp p_{12}})\!=\!
X^{(J)}_{\nu_1...\nu_{J}}(k_3^{\perp p_{12}})
 O^{\nu_1...\nu_{J}}_{\nu'_1...\nu'_{J}}(\perp p_{12}) \nn \\
&&
\times\biggl[\int\limits^\infty_{(m_1+m_2)^2}\frac{ds'_{12}
\langle\lam(s'_{12},s'_{13},s'_{23})\rangle^{(J)}_{12}}{\pi(s'_{12}-s_{12}-i0)}
 \int
X^{(J)}_{\nu'_1 ...\nu'_J} (k'^{\perp p_{12}}_{12})
 d\Phi_{12} (p'_{12};k'_1\, , k'_2)
X^{(J)}_{\nu''_1 ...\nu''_J} (k'^{\perp p_{12}}_{12}) G_J^L(s'_{12})
\biggr ]
\nn \\
&&\times
 O^{\nu''_1...\nu''_{J}}_{\mu_1...\mu_{J}}(\perp
p_{12}) X^{(J)}_{\mu_1...\mu_{J}}(k_{12}^{\perp p_{12}}).
 \end{eqnarray}
 Recall that $\langle\lam(s'_{12},s'_{13},s'_{23})\rangle^{(J)}_{12}$
depends on $s'_{12}$ only. Indeed, the projection
$\langle\lam(s_{12},s_{13},s_{23})\rangle^{(J)}_{12}$ is determined
by the following expansion of the non-singular term:
\be
\label{Tu30c}
\lam(s_{12},s_{13},s_{23})=\sum_{J'}
X^{(J')}_{\nu_1...\nu_{J'}}(k^{\perp p_{12}}_{3})
\langle\lam(s_{12}),s_{13},s_{23}\rangle^{(J')}_{12}
X^{(J')}_{\mu_1...\mu_{J'}}(k_{12}^{\perp p_{12}}), \nn \\
\ee
so we have
\bea \label{Tu30d}
&&\langle\lam(s_{12},s_{13},s_{23})\rangle^{(J)}_{12} =
\int d\Phi_{h_{in}3} (p_{12};P,-k_3)
X^{(J)}_{\nu_1...\nu_{J}}(k^{\perp p_{12}}_{3})
             \lam(s_{12},s_{13},s_{23})
\nn \\ &&\times
 X^{(J)}_{\mu_1...\mu_{J'}}(k_{12}^{\perp p_{12}})
d\Phi_{12}(p_{12};k_1,k_2) \!\int\!\! d\Phi_{h_{in}3}
(p_{12};P,-k_3) \biggl(\!X^{(J)}_{\nu'_1...\nu'_{J}}(k^{\perp
p_{12}}_{3})\!\biggr)^2 \nn \\
&&\times\int d\Phi_{12} (p_{12};k_1,k_2)
\biggl(X^{(J)}_{\mu'_1...\mu'_{J}}(k^{\perp p_{12}}_{12})\biggr)^2 \ .
\ee
In the integrand  (\ref{Tu30b}) the energy squared is $s'_{12}$.
Hence, we should use
$\langle\lam(s'_{12},s'_{13},s'_{23})\rangle^{(0)}_{12}$
in the calculation.

\begin{figure}[h]
\centerline{\epsfig{file=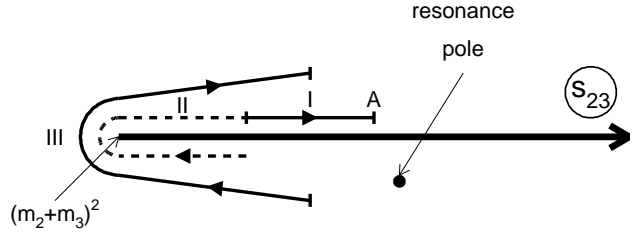, width=0.5\textwidth}}
\caption{\label{VcontourC}
The integration contour $C(s'_{12})$.}
 \end{figure}

Likewise, we calculate the amplitudes of processes of Fig.
\ref{Tcha6-8}c,d. As a result, we have the equation:
\be \label{Tu31a}
&&\hspace{-7mm}A_{12}^{(J-J)}(s_{12},s_{13},s_{23})=
 X^{(J)}_{\mu_1...\mu_{J}}(k^{\perp p_{12}}_{3})
 \nn\\
 &&\times
\biggl (B^{(J)}_{\lambda -12}(s_{12})+B^{(J)}_{13 -12}(s_{12})+
B^{(J)}_{23 -12}(s_{12}) \biggr )
\frac{X^{(J)}_{\mu_1...\mu_{J}}(k_{12}^{\perp p_{12}})}
{1-B^{(J)}_{12}(s_{12})} .~~~~
\ee
Here
\be
B^{(J)}_{i3 -12}(s_{12})=\hspace{-3mm}
  \int\limits_{(m_1+m_2)^2}^{\infty}\hspace{-3mm}
 \frac{ds'_{12}}{\pi}
 \langle A_{i3}^{(J-J)}(s'_{12},s'_{13},s'_{23})\rangle^{(J)}_{12}
\frac{\rho^{(J)}_{12}(s'_{12})} {s'_{12}- s_{12}-i0}
G_0^L(s'_{12})
\label{Tu31b}.\nn \\
\ee
Let us stress once~more~that~in~(\ref{Tu31b})~the~terms
$\langle A^{(J-J)}_{13}(s'_{12},s'_{13},s'_{23})\rangle^{(J')}_{12}$ and
$\langle A^{(J-J)}_{23}(s'_{12},s'_{13},s'_{23})\rangle^{(J')}_{12}$
 depend on $s'_{12}$ only. This is the result of the following expansions:
\begin{eqnarray} \label{Tu31c}
A^{(J-J)}_{13}(s'_{12},s'_{13},s'_{23})&=&\sum_{J'}
X^{(J')}_{\nu_1...\nu_{J'}}(k'^{\perp p_{12}}_{3})
\langle A^{(J-J)}_{13}(s'_{12},s'_{13},s'_{23})\rangle^{(J')}_{12}
X^{(J')}_{\mu_1...\mu_{J'}}(k'^{\perp p'_{12}}_{12}), \nn \\
A^{(J-J)}_{23}(s'_{12},s'_{13},s'_{23})&=&\sum_{J'}
X^{(J')}_{\nu_1...\nu_{J'}}(k'^{\perp p_{12}}_{3})
\langle A^{(J-J)}_{23}(s'_{12},s'_{13},s'_{23})\rangle^{(J')}_{12}
X^{(J')}_{\mu_1...\mu_{J'}}(k'^{\perp p'_{12}}_{12}).
\end{eqnarray}
 The integration over the phase space in the calculations of
$A^{(J-J)}_{i3}(s'_{12},s'_{13},s'_{23})$ is performed in a way
analogous to that for $J=0$. The contour integrals read (see Fig.
\ref{VcontourC}) :
\be
 \langle A^{(J-J)}_{i3}(s'_{12},s'_{13},s'_{23})\rangle^{(J)}_{i3}
&=& \int\limits_{-1}^{+1}
 \frac{dz'}{2}A^{(J-J)}_{i3}(s'_{12},s'_{13},s'_{23})
\\ &\equiv&\!
 \int\limits_{C_{i3}(s'_{12})}\! \frac{ds_{i3}}{ 4|\vk_i||\vk_3|}
 A^{(J-J)}_{i3}(s'_{12},s'_{13},s'_{23}),\quad i=1,2.\nn
\label{Tu31d}
\ee
The definition of the contours $C_i(s_{12})$  is
given in (\ref{Tu15a}).

\subsubsection{
Dispersion relation equations for a three-body system with resonance
interaction  in the two-particle states of the outgoing hadrons}

In this section we  consider the case when the outgoing particles
interact  due to two-particle resonances. Such a situation occurs,
for example, in the reaction
$p\bar p ( at\,  rest,\, level\,^1S_0)\to \pi\pi\pi$:
in the $0^{++}$-wave of the pion--pion
amplitude, there is a set of comparatively narrow resonances while a
non-resonance background can be described as a broad resonance.
Another possibility to introduce the background contribution in this
model is to add pole (resonance) terms beyond (for example, above)
the region of application of the amplitude.

In order to avoid cumbersome formulas, we consider, as before, a
reaction  of the type $h( ^1S_0)\to P_1P_2P_3 $, with the $S$-wave
interactions of the outgoing pseudoscalars $P_iP_j\to P_iP_j$.

{\centerline{\bf (i) Two-particle resonance amplitude.}}

We start with the dispersion representation of the two-particle
amplitude for this particular case. The first resonance term (the
one-fold scattering block) of the amplitude $P_iP_j\to P_iP_j$ can
be written in the form
\be \label{Td-r1}
\sum\limits_{n}
\frac{g_{ij}^{(n)2}(s_{ij})}{M^2_n- s_{ij}}\ ,
\ee
where $M_n$ is a
non-physical mass of the $n$-resonance, and vertex
$g_{ij}^{(n)}(s_{ij})$ describes its decay into two particles
$P_iP_j$. Experimental data tell us that vertices
$g_{ij}^{(n)}(s_{ij})$ can be successfully approximated by the
 energy dependence: $g_{ij}^{(n)}(s_{ij})\sim \exp(-s_{ij}/\mu^2)$
with the universal slope $\mu^2\simeq 0.5$ GeV$^2$.
 Below we assume this universality:
 \be \label{Td-r2}
g_{ij}^{(n)}(s_{ij}) = g_{ij}^{(n)}\, f(s_{ij})
\ee
 where $g_{ij}^{(n)}$ is a constant and $f(s_{ij})$ is a universal
form factor of the type $\exp(-s_{ij}/\mu^2)$. If so, the two-fold
scattering term of the amplitude contains the universal loop diagram
$b(s_{ij})$:
 \begin{eqnarray} \label{Td-r3}
&&\hspace{-2mm}
A^{(0)}_{(P_iP_j\to P_iP_j)_{\rm two-fold}}(s_{ij})=
f(s_{ij})\sum\limits_{n}
 \frac{g_{ij}^{(n)2}}{M^2_n- s_{ij}}\, b(s_{ij}) \sum\limits_{n'}
  \frac{g_{ij}^{(n')2}}{M^2_{n'}- s_{ij}}\, f(s_{ij}),\nonumber \\
&&
  b(s_{ij})
=  \int\limits_{(m_i+m_j)^2}^{\infty}
 \frac{ds'}{\pi}
\frac{\rho^{(0)}_{ij}(s')f^2(s') }{s' - s_{ij}-i0}.
\end{eqnarray}
 Summing up the terms with different numbers of loops, one obtains
 the following expression for the amplitude:
\be
A^{(0)}_{(P_iP_j\to P_iP_j)}(s_{ij})=
\frac{f^2(s_{ij})\sum\limits_{n} \frac{g^{(n)2}_{ij}}{M^2_n- s_{ij}}}
{1-b(s_{ij})\sum\limits_{n'} \frac{g^{(n')2}_{ij}}{M^2_{n'}- s_{ij}}}\; .
\label{Tf1}
\ee
 Since the loop diagram has the following real and imaginary parts:
\be
b(s_{ij})&=& i\rho^{(0)}_{ij}(s_{ij})f^2(s_{ij}) +
 P \int\limits_{(m_i+m_j)^2}^{\infty}
 \frac{ds'}{\pi}
\frac{\rho^{(0)}_{ij}(s')f^2(s') }{s' - s_{ij}}
\nn \\&=&i{\rm Im}\, b(s_{ij}) + {\rm Re}\, b(s_{ij})       \; ,
\label{Tf1_2}
\ee
 the scattering amplitude (\ref{Tf1}) can easily be rewritten in the
$K$-matrix form for the case when an $S$-wave state contains several
resonances.

{\centerline {\bf (ii) Three particle production amplitude
             $h( ^1S_0)\to P_1P_2P_3 $. }}

The decay amplitude is given by an equation of the type of
(\ref{Tu4}). In the term $\lam(s_{12},s_{13},s_{23})$, however, we
should take into account the prompt production of resonances; it is
convenient to consider their widths as well. Correspondingly, we
replace
\begin{eqnarray} \label{Tr1}
&&\lam(s_{12},s_{13},s_{23})\to
\Lambda_{12}(s,s_{12})+ \Lambda_{13}(s,s_{13})+\Lambda_{23}(s,s_{23}),
\\
&&  \Lambda_{ij}(s,s_{ij}) =
\sum\limits_{n}\Lambda^{(n)}_{ij}(s,s_{ij})
\frac{g^{(n)}_{ij}}{M^2_n- s_{ij}}f(s_{ij})
\frac{1}
{1-b(s_{ij})\sum\limits_{n'} \frac{g^{(n')2}_{ij}}{M^2_{n'}- s_{ij}}}\;
,  \nonumber
\end{eqnarray}
and write for the full amplitude:
\begin{eqnarray} \label{Tr2}
\hspace{-9mm}A_{h( ^1S_0)\to P_1P_2P_3}(s_{12},s_{13},s_{23})&=&
\Lambda_{12}(s,s_{12})\!+\! \Lambda_{13}(s,s_{13})\!+\!\Lambda_{23}(s,s_{23})
\nonumber \\
&+& A_{12}(s,s_{12})\! +\!
A_{13}(s,s_{13})\! +\!A_{23}(s,s_{23}),
\end{eqnarray}
 where the terms $A_{ij}(s,s_{ij})$ describe processes with interactions
of all particles, $P_1$, $P_2$ and $P_3$.  The term $A_{12}(s,s_{12})$
reads:
\begin{eqnarray}  \label{Tr3}
 A_{12}(s,s_{12})&=&\biggl
(B_{13-12}(s,s_{12})+B_{23-12}(s,s_{12})\biggr)
\nn \\ &\times&
\sum\limits_{n}\frac{ g^{(n)2}_{12}}{M^2_n- s_{12}} f(s_{12})
 \frac{1}
{1-b(s_{12})\sum\limits_{n'} \frac{g^{(n')2}_{12}}{M^2_{n'}- s_{12}}},
\\
B_{i3 -12}(s_{12})&=&
  \int\limits_{(m_1+m_2)^2}^{\infty}
 \frac{ds'_{12}}{\pi}
 \langle \Lambda_{i3}(s,s'_{i3})+ A_{i3}(s,s'_{i3})
\rangle^{(0)}_{12} \;
\frac{\rho^{(0)}_{12}(s'_{12})f(s'_{12})} {s'_{12}- s_{12}-i0}
\, . \nn
\end{eqnarray}
Analogous relations for $ A_{13}(s,s_{13})$ and $ A_{23}(s,s_{23})$
give us three non-homogenous equations for three amplitudes thus
solving in principle the problem of construction of the three-body
amplitude under the constraints of analyticity and unitarity.

The equations written here require a comment. We realise the
convergence of the loop diagrams with the help of cutting vertices
or, what is the same, the universal form factor. The convergence of
a loop diagram can be realised in other ways as well. For example,
in \cite{TAA-eta,TAA-3mesons} a special cutting function was
introduced into the integrands; one may use the subtraction
procedure as it is done in \cite{TGribov-K2,TAA}. The technical
variations are of no importance, the only essential point is that
for the convergence of the considered diagrams we have to introduce
additional parameters -- here it is the form factor slope $\mu^2$,
see Eq. (\ref{Td-r2}) and the corresponding discussion.

{\it Miniconclusion}

In this Appendix we have presented some characteristic features of
the spectral integral equations  for three-body systems. The
technique can be used both for the determination of levels of
compound systems and their wave functions (for instance, in the
method of an accounting  of the leading singularities
\cite{TAA-He-3} -- this method was applied to the three-nucleon
systems, $H_3$ and $He_3$, and for determination of analytical
properties of multiparticle production amplitudes when the produced
resonances are studied.

We do not present here formulas for the reactions we have studied
--- the formulas are rather cumbersome. An example can be found, as it
was mention above, in \cite{TAA-3mesons} where a set of equations
for reactions $p\bar p \to \pi\pi\pi,\pi\eta\eta,\pi K\bar K$ was
written.

\begin{figure}
\centerline{\epsfig{file=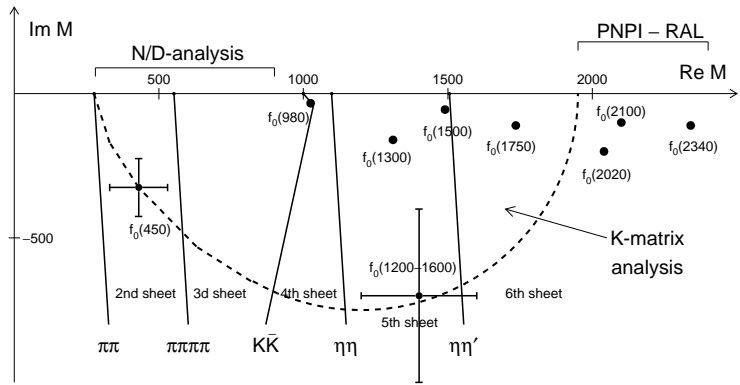,width=10cm}}
\caption{Complex-$M$ plane for the $(IJ^{PC}=00^{++})$ mesons. The
dashed line encircles the part of the plane where the $K$-matrix
analysis \protect\cite{Tkm,TYF,Tepja} reconstructs the analytical $K$-matrix
amplitude: in this area the poles corresponding to resonances
$f_0(980)$, $f_0(1300)$, $f_0(1500)$, $f_0(1750)$ and the broad
state $f_0(1200-1600)$ are located. Beyond this area, in the
low-mass region, the pole of the light $\sigma$-meson is located
(shown by the point the position of pole, $M=(430-i320)$ MeV,
corresponds to the result of $N/D$ analysis ; the crossed bars stand
for $\sigma$-meson pole found in \protect\cite{Tufn,N/D}). In the
high-mass region one has resonances $f_0(2030),f_0(2100),f_0(2340)$
\protect\cite{Tpnpi-ral}.  Solid lines stand for the cuts related to the
thresholds $\pi\pi,\pi\pi\pi\pi,K\bar K,\eta\eta,\eta\eta'$.}
\label{Xpolesco2}
\end{figure}

\begin{figure}[h]
\centerline{\epsfig{file=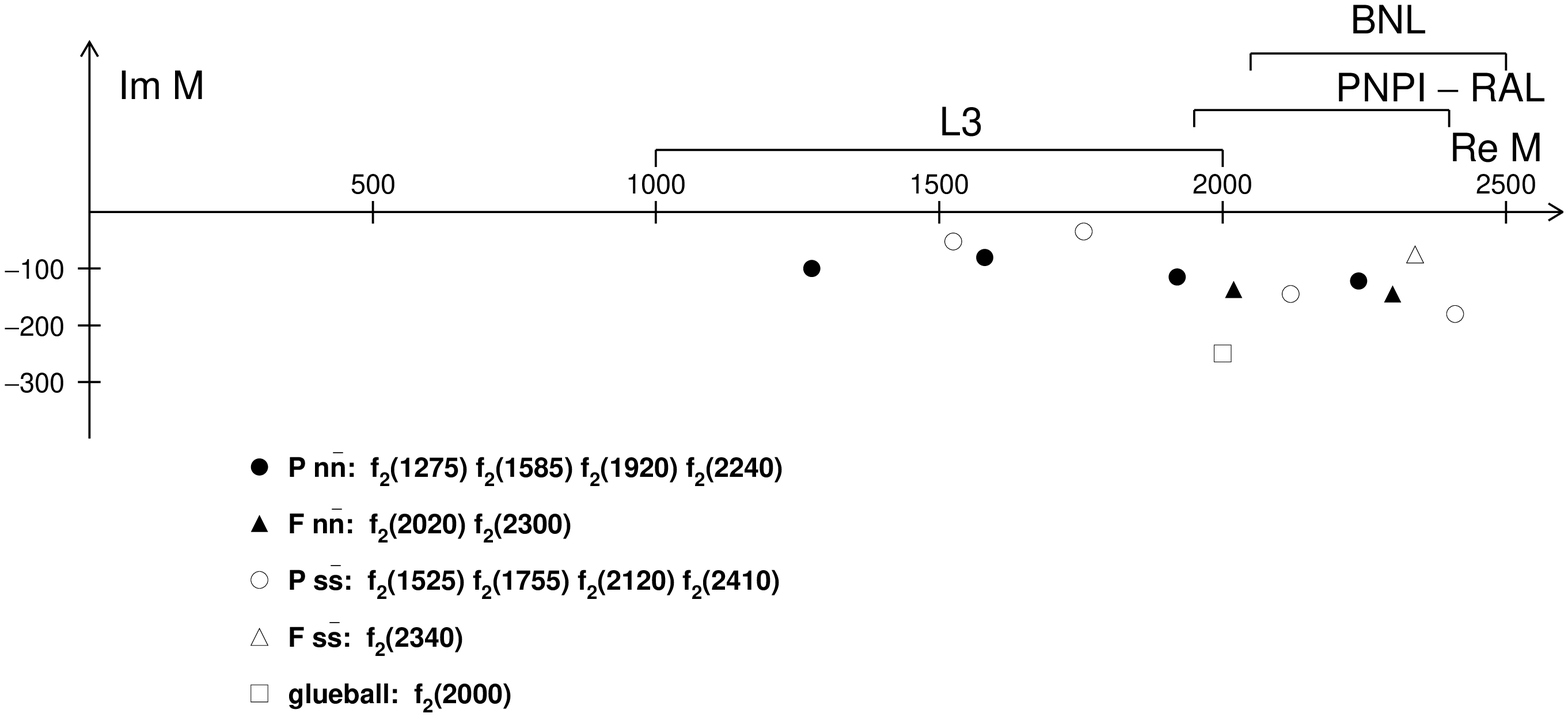,width=0.98\textwidth}}
\caption{Position of the $f_2$-poles on the complex-$M$ plane:
states with dominant $^3P_2n\bar n$-component (full circle),
$^3F_2n\bar n$-component (full triangle), $^3P_2s\bar s$-component
(open circle), $^3F_2s\bar s$-component (open triangle); the
position of the tensor glueball is shown by the open square. Mass
regions studied by the groups L3 \protect\cite{L3}, PNPI-RAL
\protect\cite{Tpnpi-ral,glueball2,AS,AMNS} and BNL \protect\cite{Tbnl,LL} are shown.}
\label{Xp-pos}
\end{figure}

The presented technique may be especially convenient for the study
of low-mass singularities in  multiparticle production amplitudes.
The long-lasting discussions on the sigma-meson
observation, (see \cite{Tsigma04,TbuggPhysRep,Tsigma05} and references
therein) indicate that this is a problem of current interest.

\section{Assignment of Mesons to Nonets}

 We present
here the assignment of the  mesons to $q\bar q$ nonets following
\cite{book3}.

In Figs. \ref{Xpolesco2} and \ref{Xp-pos} we show the
positions of the $(IJ^{PC}=00^{++})$ and $(IJ^{PC}=02^{++})$
resonances in the complex-$M$ planes found in the previous analyses,
see \cite{book3} for details.

In  Eq. (\ref{D1}) we collected all considered meson $q\bar q$
states in nonets according to their SU(3)$_{flavour}$
attribution
(the $^{2S+1}L_J$ states of $q \bar{q}$ mesons with n =
1, 2, 3 and 4). The singlet and octet states, with the
same values of the total angular momentum, are mixed.
\begin{equation} \label{D1}
{\footnotesize
  {\begin{tabular}{c|cccc|cccc}
    & \multicolumn{4}{c|}{n=1} & \multicolumn{4}{c}{n=2} \\
    \raisebox{1ex}{$q \bar{q}$-mesons}& I=1 & I=0 & I=0 & I=$\frac{1}{2}$ & I=1 & I=0 & I=0 & I=$\frac{1}{2}$ \\ [1.5ex]
    \hline
    & \multicolumn{4}{c|}{} & \multicolumn{4}{c}{} \\ [-2ex]
    $^1S_0(0^{-+})$ & $\pi(140)$ & $\eta(547)$ & $\eta'(958)$ & $K(500)$ & $\pi(1300) $ & $\eta(1295)$ & $\eta(1410)$ & $K(1460)$ \\
    $^3S_1(1^{--})$ & $\rho(775)$ & $\omega(782)$ & $\phi(1020)$ & $K^\ast$(890) & $\rho(1460)$ & $\omega(1430)$ & $\phi(1650)$ & \\ [0.5ex]
    \hline
    & \multicolumn{4}{c|}{} & \multicolumn{4}{c}{} \\ [-2ex]
    $^1P_1(1^{+-})$ & $b_1(1229)$ & $h_1(1170)$ & $h_1(1440)$ & $K_1(1270)$ & $ b_1(1620)$ & $ h_1(1595)$ & ${\bf h_1(1790)}$ & $K_1(1650)$ \\
    $^3P_0(0^{++}$ & $a_0(980)$ & $f_0(980)$ & $f_0(1300)$ & $K_0(1425)$ & $a_0(1474)$ & $f_0(1500)$ & $f_0(1750)$ & $K_0(1820)$ \\
    $^3P_1(1^{++})$ & $a_1(1230)$ & $f_1(1282)$ & $f_1(1426)$ & $K_1(1400)$ & $a_1(1640)$ &$f_1(1518)$ &${\bf f_1(1780)}$ & \\
    $^3P_2(2^{++})$ & $a_2(1320)$ & $f_2(1275)$ & $f_2(1525)$ & $K_2(1430)$ & $a_2(1675)$ & $f_2(1565)$ & $f_2(1755)$ & $K_2(1980)$ \\ [0.5ex]
    \hline
    & \multicolumn{4}{c|}{} & \multicolumn{4}{c}{} \\ [-2ex]
    $^1D_2(2^{-+})$ &$\pi_2(1676)$&$\eta_2(1645)$&$\eta_2(1850)$&$K_2(1800)$&$\pi_2(2005)$&$\eta_2(2030)$&${\bf eta_2(2150)}$& \\
    $^3D_1(1^{--})$ &$\rho(1700)$&$\omega(1670)$&&$K_1(1680)$&$\rho(1970)$&$\omega(1960)$&& \\
    $^3D_2(2^{--})$ &$\rho_2(1940)$&$\omega_2(1975)$&&$K_2(1580)$&$\rho_2(2240)$&$\omega_2(2195)$&&$ K_2(1773)$ \\
    $^3D_3(3^{--})$ &$\rho_3(1690)$&$\omega_3(1667)$&$\phi_3(1854)$&$K_3(1780)$&$\rho_3(1980)$&$\omega_3(1945)$&${\bf \phi _3(2140)}$& \\       \\ [0.5ex]
    \hline
    & \multicolumn{4}{c|}{} & \multicolumn{4}{c}{} \\ [-2ex]
    $^1F_3(3^{+-})$ &$b_3(2032)$&$h_3(2025)$&&&$ b_3(2245)$&$h_3(2275)$&& \\
    $^3F_2(2^{++})$ &$a_2(2030)$&$f_2(2020)$&$f_2(2340)$&&$a_2(2255)$&$f_2(2300)$&${\bf f_2(2570)}$& \\
    $^3F_3(3^{++})$ &$a_3(2030)$&$f_3(2050)$&&$K_3(2320)$&$a_3(2275)$&$f_3(2303)$&& \\
    $^3F_4(4^{++})$ &$a_4(2005)$&$f_4(2025)$&${\bf f_4(2100)}$&$K_4(2045)$&$a_4(2255)$&${\bf f_4(2150)}$&$f_4(2300)$& \\ [0.5ex]
    \hline
    & \multicolumn{4}{c|}{} & \multicolumn{4}{c}{} \\ [-2ex]
    $^1G_4(4^{-+})$ &$\pi_4(2250)$&$\eta_4(2328)$&&$K_4(2500)$&&&& \\
    $^3G_3(3^{--})$ &$\rho_3(2240)$&&&&${\bf \rho_3(2510)}$&&& \\
    $^3G_4(4^{--})$ &&&&&&&& \\
    $^3G_5(5^{--})$ &$\rho_5(2300)$&&&$K_5(2380)$&${\bf \rho_5(2570)}$&&& \\    [0.5ex]
    \hline
    & \multicolumn{4}{c|}{} & \multicolumn{4}{c}{} \\ [-2ex]
    $^3H_6(6^{++})$ &$a_6(2450)$&$f_6(2420)$&&&&&& \\   [0.5ex]
    \hline
\hline
& \multicolumn{4}{c|}{n=3} & \multicolumn{4}{c}{n=4}  \\
\raisebox{1ex}{$q \bar{q}$-mesons}& I=1 & I=0 & I=0 & I=$\frac{1}{2}$ & I=1 & I=0 & I=0 & I=$\frac{1}{2}$   \\ [1.5ex]
\hline
& \multicolumn{4}{c|}{} & \multicolumn{4}{c}{}  \\ [-2ex]
$^1S_0(0^{-+})\hspace{-2mm}$&$\pi(1800)\hspace{-2mm}$ & $ \eta(1760)\hspace{-2mm}$ & $ {\bf \eta (1880)\hspace{-2mm}}$ & $K(1830)\hspace{-2mm}$ & $\pi(2070)\hspace{-2mm}$ & $\eta(2010)\hspace{-2mm}$ &$\eta(2190)\hspace{-2mm}$ &  \\
$^3S_1(1^{--})\hspace{-2mm}$ &$\rho(1870)\hspace{-2mm}$&${\bf\omega(1830)\hspace{-2mm}}$&${\bf \phi(1970)\hspace{-2mm}}$& &$\rho(2110)\hspace{-2mm}$&$\omega(2205)\hspace{-2mm}$&${\bf\phi (2300)\hspace{-2mm}}$& \\ [0.5ex]
\hline
& \multicolumn{4}{c|}{} & \multicolumn{4}{c}{}   \\ [-2ex]
$^1P_1(1^{+-})\hspace{-2mm}$ & $b_1(1960)\hspace{-2mm}$ & $h_1(1965)\hspace{-2mm}$ & ${\bf h_1(2090)\hspace{-2mm}}$& &$b_1(2240)\hspace{-2mm}$ & $h_1(2215)\hspace{-2mm}$ &  & \\
$^3P_0(0^{++})\hspace{-2mm}$ & ${\bf a_0(1780)\hspace{-2mm}}$ & $f_0(2040)\hspace{-2mm}$ & $f_0(2105)\hspace{-2mm}$ && $ a_0(2025)\hspace{-2mm}$ & $ f_0(2210)\hspace{-2mm}$ & $f_0(2340)\hspace{-2mm}$ & \\
$^3P_1(1^{++})\hspace{-2mm}$ & $a_1(1930)\hspace{-2mm}$ &$f_1(1970)\hspace{-2mm}$ &${\bf f_1(2060)\hspace{-2mm}}$ & & $a_1(2270)\hspace{-2mm}$ & ${\bf f_1(2214)\hspace{-2mm}}$ & $ f_1(2310)\hspace{-2mm}$ & \\
$^3P_2(2^{++})\hspace{-2mm}$ & $a_2(1950)\hspace{-2mm}$ & $f_2(1920)\hspace{-2mm}$ & $f_2(2120)\hspace{-2mm}$ &  & $a_2(2175)\hspace{-2mm}$ & $f_2(2240)\hspace{-2mm}$ & $ f_2(2410)\hspace{-2mm}$ & \\ [0.5ex]
\hline
& \multicolumn{4}{c|}{} & \multicolumn{4}{c}{}  \\ [-2ex]
$^1D_2(2^{-+})\hspace{-2mm}$ &$\pi_2(2245)\hspace{-2mm}$&$\eta_2(2248)\hspace{-2mm}$& $ {\bf \eta_2 (2380)\hspace{-2mm}}$& && $ {\bf \eta_2 (2520)\hspace{-2mm}}$&& \\
$^3D_1(1^{--})\hspace{-2mm}$ &$\rho(2265)\hspace{-2mm}$&$\omega(2330)\hspace{-2mm}$&&&&&& \\
$^3D_2(2^{--})\hspace{-2mm}$ &&&&$ K_2(2250)\hspace{-2mm}$&&&& \\
$^3D_3(3^{--})\hspace{-2mm}$ &$\rho_3(2300)\hspace{-2mm}$&$\omega_3(2285)\hspace{-2mm}$&$ {\bf \phi _3(2400)\hspace{-2mm}}$&&&&& \\ [0.5ex]
\end{tabular}  }
} \\
\end{equation}

In
the lightest nonets we can determine mixing angles more or less
reliably, but for the higher excitations the estimates of the mixing
angles are very ambiguous. In addition, isoscalar states can contain
significant glueball components. For
these reasons, we give only the nonet ($9=1\oplus8$) classification
of mesons. States that are predicted but not yet reliably
established are shown in boldface.

\section*{Acknowledgments}

We thank A.V. Anisovich, L.G. Dakhno, J. Nyiri, V.A. Nikonov, M.A. Matveev
for helpful discussions. The paper was supported by the RFFI grant
07-02-01196-a.

 \end{document}